\documentclass{aa}  

\usepackage{graphics,graphicx}
\usepackage{xcolor}
\usepackage{enumitem}
\usepackage{caption}

\usepackage{txfonts}
\usepackage{longtable}
\usepackage[unicode=true,pdfusetitle,
 bookmarks=true,bookmarksnumbered=false,bookmarksopen=false,
 breaklinks=false,pdfborder={0 0 0},pdfborderstyle={},backref=false,colorlinks=true]
 {hyperref}
\hypersetup{
 urlcolor=blue,citecolor=blue}

\DeclareMathOperator{\sech}{sech}
\begin{document}

   \title{SRG/eROSITA 3D mapping of the ISM using X-ray absorption spectroscopy}

   \author{E. Gatuzz\inst{1}, 
           J. Wilms\inst{2}, 
           A. Zainab\inst{2}, 
           S. Freund\inst{1}, 
           P. C. Schneider\inst{3}, 
           J. Robrade\inst{3}, 
           S. Czesla\inst{3}, \\  
           J. A. Garc\'ia \inst{4} \and
           T. R. Kallman\inst{4}    
          }

   \institute{Max-Planck-Institut f\"ur extraterrestrische Physik, Gie{\ss}enbachstra{\ss}e 1, 85748 Garching, Germany\\
              \email{egatuzz@mpe.mpg.de}
         \and
             Dr. Karl Remeis-Observatory \& ECAP, Friedrich-Alexander-Universit\"at Erlangen-N\"urnberg, Sternwartstr. 7, 96049 Bamberg, Germany
         \and
             Hamburger Sternwarte, Universit\"at Hamburg, Gojenbergsweg 112, 21029 Hamburg, Germany 
         \and
             NASA Goddard Space Flight Center, Greenbelt, MD 20771, USA
             }
 
   \date{Received XXX; accepted YYY}

  \abstract  
{
We present a detailed study of the hydrogen density distribution in the local interstellar medium (ISM) using the X-ray absorption technique.
Hydrogen column densities were precisely measured by fitting X-ray spectra from coronal sources observed during the initial {\it eROSITA} all-sky survey (eRASS1).
Accurate distance measurements were obtained through cross-matching Galactic sources with the {\it Gaia} third data release (DR3).
Despite the absence of a discernible correlation between column densities and distances or Galactic longitude, a robust correlation with Galactic latitude was identified. 
This suggests a decrease in ISM material density along the vertical direction away from the Galactic plane. 
To further investigate, we employed multiple density laws to fit the measured column densities, revealing constraints on height scale values ($8 < h_{z} < 30$~pc). 
Unfortunately, radial scales and central density remain unconstrained due to the scarcity of sources near the Galactic center. 
Subsequently, a 3D density map of the ISM was computed using a Gaussian processing approach, inferring hydrogen density distribution from hydrogen column densities. 
The results unveiled the presence of multiple beams and clouds of various sizes, indicative of small-scale structures. 
Large density regions were identified at approximately 100~pc, consistent with findings in dust reddening studies, potentially associated with the Galactic Perseus arm.
Moreover, high-density regions were pinpointed in proximity to the Orion star-forming region and the Chamaeleon molecular complex, enriching our understanding of the intricate structure of the local interstellar medium. 
}
 
    \titlerunning{eROSITA mapping of the ISM X-ray absorption}
    \authorrunning{Gatuzz et al.}
   \maketitle


\section{Introduction}\label{sec_int}
The interstellar medium (ISM) is one of the most essential constituents of galaxies, affecting the stellar formation and evolutionary processes.
This environment includes cold (<10$^{4}$ K), warm (10$^{4}$ -- 10$^{6}$ K) and hot (>10$^{6}$ K) components \citep[see][and references therein]{dra11}.
The cold component, in particular, plays a vital role in the Galactic evolution given that it is dominated by Hydrogen, the most abundant chemical element in the universe \citep{san08}.
Multiple radio and H$\alpha$ surveys have shown that most of the {\rm H}~{\sc i} gas resides in a thin disc along the Galactic plane, with the Sun embedded within it \citep{dic90,har97,baj05,kal05}. 
However, they have limited angular resolution, typically $\sim$0.25 deg$^2$ or larger. 
21~cm emission measurements have been used to map the cold component on angular scales of 0.6$^{\circ}$ in combination with distances inferred from velocities and assuming certain galactic rotation curves \citep{kal05,win16}.
However, 21~cm emission measurements cover the entire galaxy and do not include the molecular component, thus making it challenging to map local spatial ISM structures.

X-ray absorption measurements constitute a powerful technique to map the ISM in detail.
By taking an X-ray source, acting as a high-energy photon lamp, and measuring the absorption spectral features (i.e., spectral edges and absorption lines), the physical conditions of the ISM can be measured.
Such analysis allows for studying fine-scale spatial structure and considering the impact of the molecular component, which also affects the shape of the spectra continuum.
Moreover, while emission measurements are sensitive to density fluctuations (via the density square dependence on emissivity), the X-ray absorption has a linear dependence of opacity on density.
Finally, X-rays generally can probe columns and distances larger than the optical measurements \citep{gat18b}.

In the last decade multiple studies have been performed to model the multiphase ISM using high-resolution X-ray spectra  \citep{sch02, tak02,jue04,jue06,yao09,lia13,pin10,pin13,gat13a,gat13b,gat14,gat15,gat16,nic16a,gat17,eck17,gat18b,gat20c,psa20,gat21}. 
\citet{gat18}, in particular, studied the distribution of the NH in the Milky Way by analyzing X-ray spectra of both Galactic and extragalactic sources.
They found a general trend of higher column densities near the Galactic plane for the cold component.
They attempted to model the gas distribution with an exponential analytical model. 
However, the lack of sources near the Galactic plane did not allow them to obtain reasonable constraints for the model parameters.
 
While it is difficult to disentangle the multiphase ISM using moderate-resolution spectra \citep[although see for counterexample][]{gat20}, CCD spectra allow to study the equivalent NH absorption for the cold component.
\citet{gat18b} computed a 3D map of the hydrogen density distribution by using NH measure from {\it XMM-Newton} Galactic sources in combination with distances from the first {\it Gaia} Data Release (DR1).
They used a Bayesian method explained in \citet{rez17} to predict the density distribution even for lines of sight with no initial observations.
They found small-scale density structures that analytic density profiles cannot model.
However, using {\it Gaia} DR1 and needing more flexibility in the modeling lead to significant uncertainties in the column densities. 
Therefore, the maps should be considered qualitatively.

In this paper, we present a 3D map of the hydrogen density distribution in the local ISM obtained from {\it eROSITA} spectral fitting in combination with distances obtained from the {\it Gaia} third data release (DR3). 
The all-sky survey conducted by {\it eROSITA} offers a unique opportunity for such analysis due to the extensive identification of X-ray Galactic sources.
The structure of the present paper is as follows.
Section~\ref{sec_dat} explains the creation of the {\it eROSITA} data sample while Section~\ref{sec_mod} describes the X-ray spectral modeling. 
Section~\ref{sec_nh} describes the NH obtained and a comparison with previous results.
The modeling of the data with density laws is shown in Section~\ref{sec_law} while the 3D map computation of the density distribution is described in Section~\ref{sec_map}.
Finally, the conclusions are summarized in Section~\ref{sec_con}. 
For the spectral analysis, we use the {\sc xspec} data fitting package (version 12.13.1\footnote{\url{https://heasarc.gsfc.nasa.gov/xanadu/xspec/}}). 
For the X-ray spectral fits, we assumed $\chi^{2}$ statistics in combination with the \citet{chu96} weighting method, which allows the analysis of data in the low counts regime providing a goodness-of-fit criterion \citep[see for example,][]{gat19a, gat20}. 
Errors are quoted at $1\sigma$ confidence level unless otherwise stated and abundances are given relative to \citet{gre98}. 
Finally, we denote the column density obtained from X-ray measurements as NH$_{X}$.

 \section{{\it eROSITA} Galactic objects sample}\label{sec_dat}
The {\it eROSITA} X-ray telescope \citep{pre21}, on board the {\it Spectrum R\"otgen Gamma} (SRG) observatory, performed the deepest all-sky survey at soft X-ray energy range ($0.2-10$~keV). 
More than 1 million sources were detected by {\it eROSITA} during the first all-sky survey (eRASS1), with $80\%$ of the sources being associated with AGNs {\bf (Merloni et al. (2024), A\&A, 682, A34)}. 
{\bf Freund et al. (2023) and Salvato et al. (2023)} performed a cross-matching between the coronal eRASS1 sources and the {\it Gaia} third data release \citep[DR3, ][]{gai21,gai23} in order to create a catalog of eRASS1 stellar sources, including {\it Gaia} distances. 
By adopting the HamStar identification procedure \citep{sch21,fre22}, they identified 137500 coronal sources with at least one optical counterpart and displayed properties of coronal X-ray sources. 
   
We produced X-ray spectra from this catalog using the {\it eROSITA} data analysis software {\tt eSASS} version 201125 with 010 processing. 
In particular, we used the {\tt srctool} task to produce source and background spectra as well as response files for each source identified in {\bf Freund et al. (2023)}.
The standard data reduction procedure includes creating circular extraction regions with radii scaled to the maximum likelihood (ML) count rate from the eRASS1 source catalog. 
Background regions are also created as annuli with sizes scaled to the ML count rate.
We combined the data from the Telescope Modules (TMs) in TM 1-4,6 and TM 5,7 because of the excess soft emission due to optical light leaking observed in TM 5 and 7 \citep{pre21}.
The last group was analyzed only for energies $>1$~keV.

\section{X-ray spectral modeling}\label{sec_mod}
Following the work done in \citet{gat18b}, we fitted each spectra with multiple models selected to represent in a phenomenological way the most commonly observed spectral shapes in astronomical sources. 
The models are (using {\sc xspec} nomenclature):

\begin{itemize}
\item Model A: An absorbed power-law model ({\sc XSPEC}: {\tt tbabs*pow}).
\item Model B: An absorbed thermal model ({\sc XSPEC}: {\tt tbabs*apec}).
\item Model C: An absorbed black-body model ({\sc XSPEC}: {\tt tbabs*bbody}).
\item Model D: An absorbed double thermal model ({\sc XSPEC}: {\tt tbabs*(apec+apec)}).
\item Model E: same as D but with free abundances for the {\tt apec} components,
\end{itemize}
 
where {\tt tbabs} is the ISM X-ray absorption model described in \citet{wil00}. 
Then, by fitting the curvature of the X-ray spectra with the models described above, we estimate NH$_{X}$ values ($X$ for X-rays). 
The best-fit parameters, the fit-statistic, and statistical uncertainties are available for each model. 
In the fitting process, first, we fixed the hydrogen column density to the value computed by \citet{wil13} and allowed the emission parameters to vary. 
Then, we keep the NH$_{X}$ as a free parameter and proceeded to compute the uncertainties.
After all models were applied, the final NH$_{X}$ value for our analysis was selected from the model for which {\tt chisqr}/d.o.f. is close to 1.0. 
It is important to note that in this study, our main interest is to obtain the most accurate emission+absorption fit to measure NH$_{X}$ instead of the physical conditions of the Galactic sources. 

\begin{figure}
\includegraphics[width=0.48\textwidth]{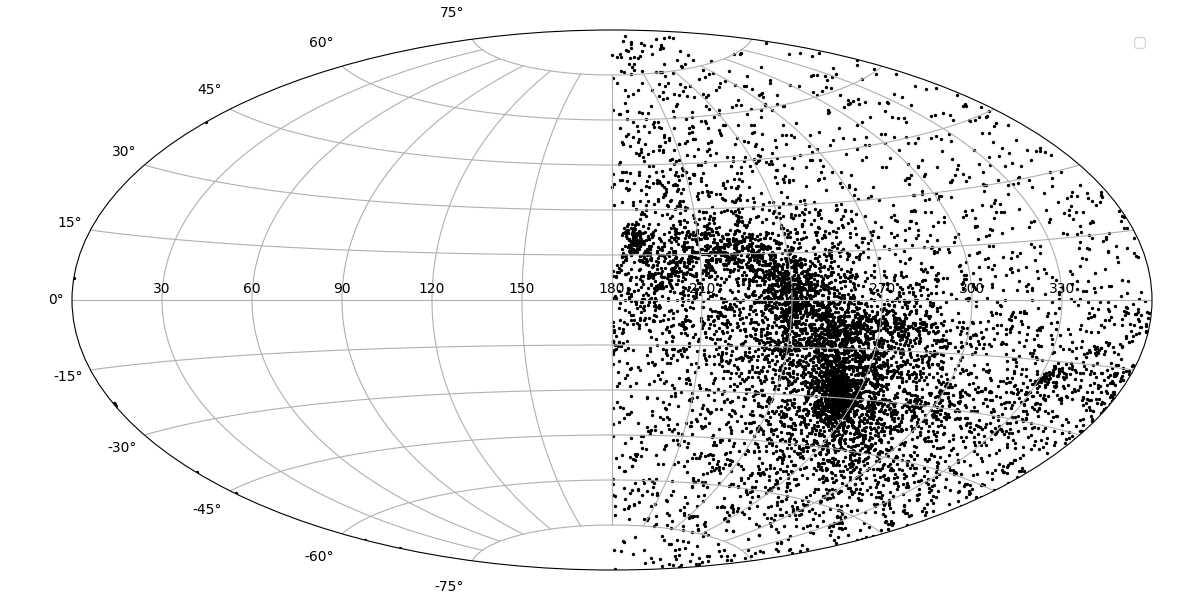}
      \caption{
      Distribution of the 8231 eRASS1 sources analyzed in this work.
      }\label{fig_ait}
   \end{figure}

\begin{figure}
\includegraphics[width=0.48\textwidth]{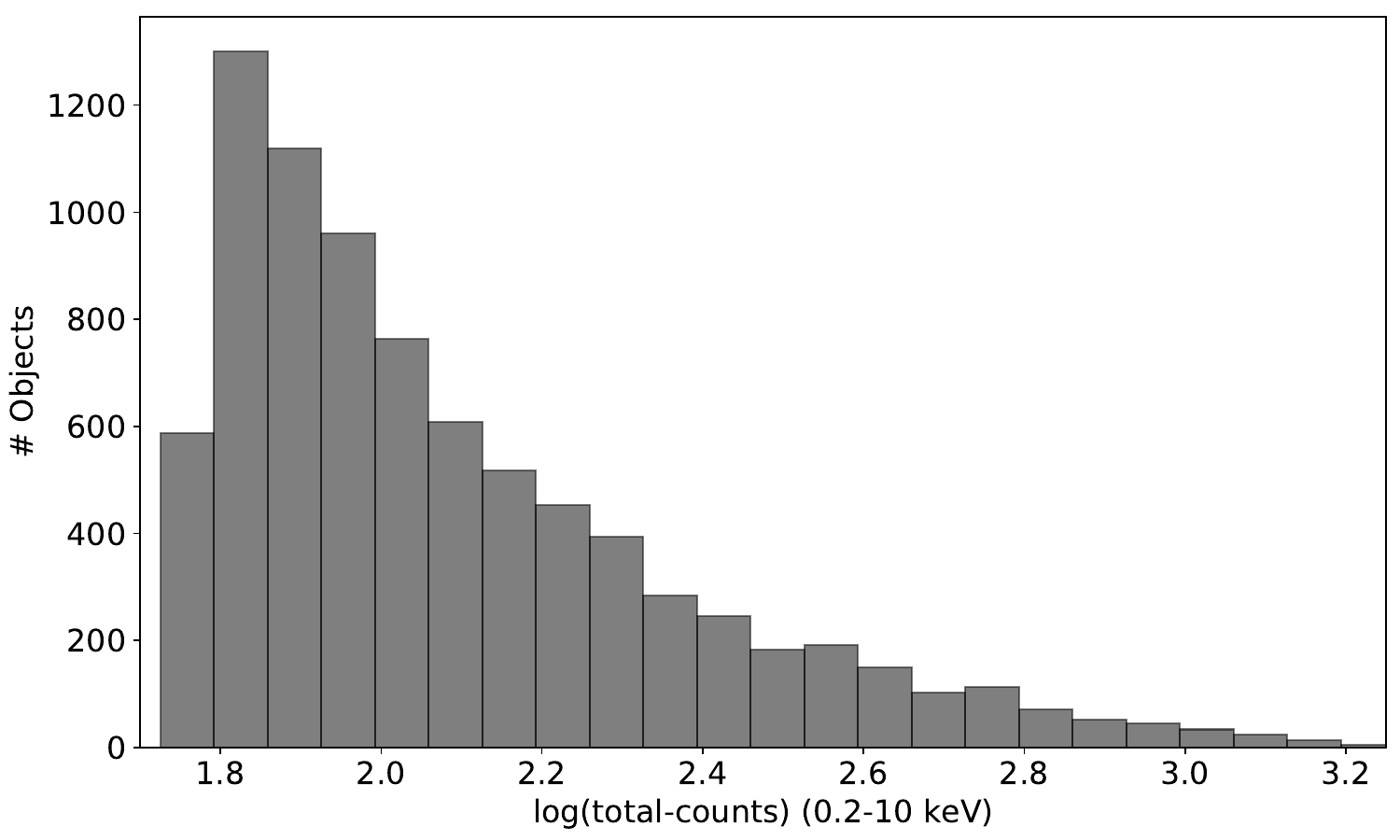}\\
\includegraphics[scale=0.35]{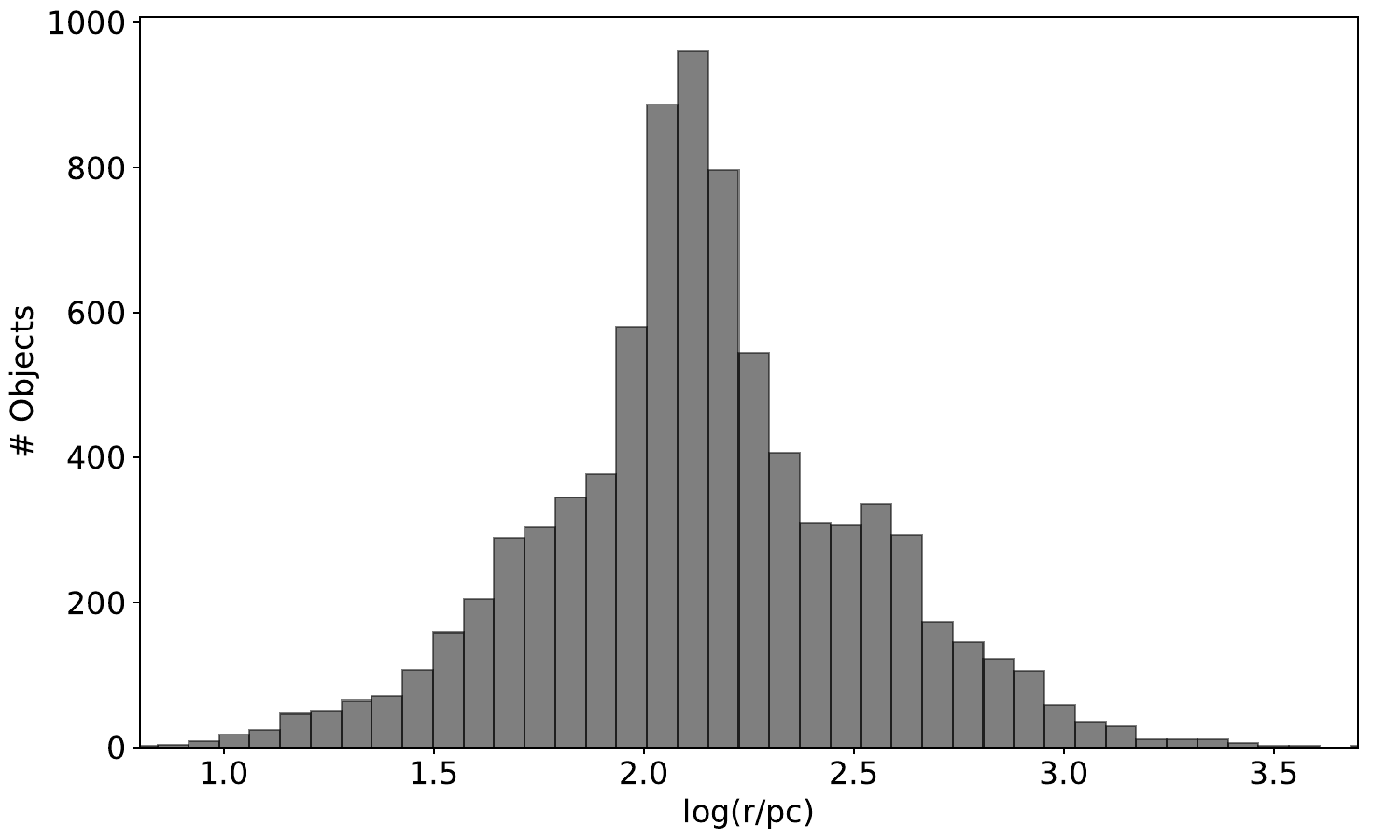}
      \caption{
  \emph{Top panel:} distribution of sources as function of the number of counts in the 0.2-10~keV energy range.
  \emph{Bottom panel:} distribution of sources as function of the distance obtained from the {\it Gaia} observatory.
      }\label{fig_his}
   \end{figure}

We obtained constrained column densities for 8231 eRASS1 sources, a sample almost one order of magnitude larger than the analysis done by \citet{gat18b}. 
We obtain only upper limits or non-realistic values (e.g., $<10^{15}$ cm$^{-2}$) for the rest of the sources, due to the poor statistics in the spectra.
We notice that in most cases, the best statistical fit is obtained for models D-E ($\sim 80\%$), which points out the complexity of the coronal spectra modeling \citep[see for example ][]{rob12,cof22}. 
Figure~\ref{fig_ait} shows the distribution of the sources in the Aitoff projection. 
Due to the SRG scanning strategy, sources near the ecliptic equator are observed six times during one day, while sources at higher ecliptic latitudes are observed for longer periods of time.
Note that the German {\it eROSITA} consortium only has access to the western hemisphere. 
Figure~\ref{fig_his} top panel shows the distribution of the sources as a function of the number of counts in the 0.2-10~keV energy range. 
As an X-ray survey, the snapshot observations lead to many sources being in the low-count regime.
Figure~\ref{fig_his} bottom panel shows the distribution of the sources as a function of the {\it Gaia} distances ($r$).
A normal distribution can be distinguished, with the peak around $125$~pc, although distances $>1000$~pc are also covered.
Such distribution indicates that our analysis of the NH$_{X}$ distribution corresponds to the local ISM.

\section{Hydrogen column densities}\label{sec_nh} 
\begin{figure*} 
\includegraphics[width=\textwidth ]{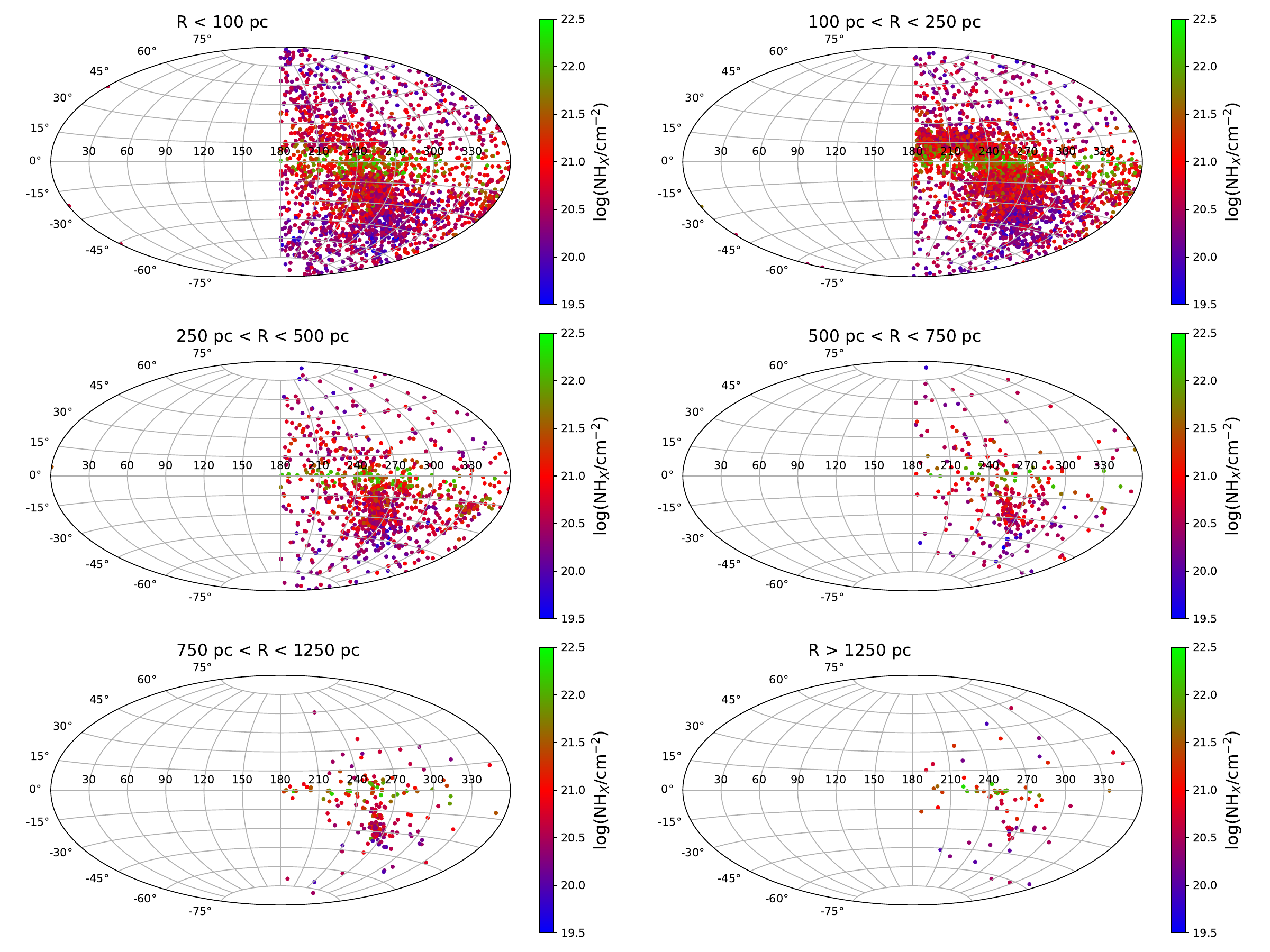}
      \caption{
      Distribution of the sources for different distance ranges in Aitoff projection.
      The colors indicate the $\log$(NH$_{X}$) obtained from the best-fit model in units of cm$^{-2}$
      }\label{fig_nh_ait}
   \end{figure*}

Figure~\ref{fig_nh_ait} shows an Aitoff projection of the Galactic sources analyzed for different distance ranges. 
The $\log$(NH$_{X}$) values obtained from the X-ray spectral fitting procedure described above are indicated by colors in units of cm$^{-2}$. 
Although most sources concentrate near the Galactic plane, the sample includes many high-latitude sources.
We have found an average column density of NH$_{X}=(0.15\pm 0.05)\times 10^{22}$ cm$^{-2}$.

\begin{figure} 
\includegraphics[width=0.48\textwidth ]{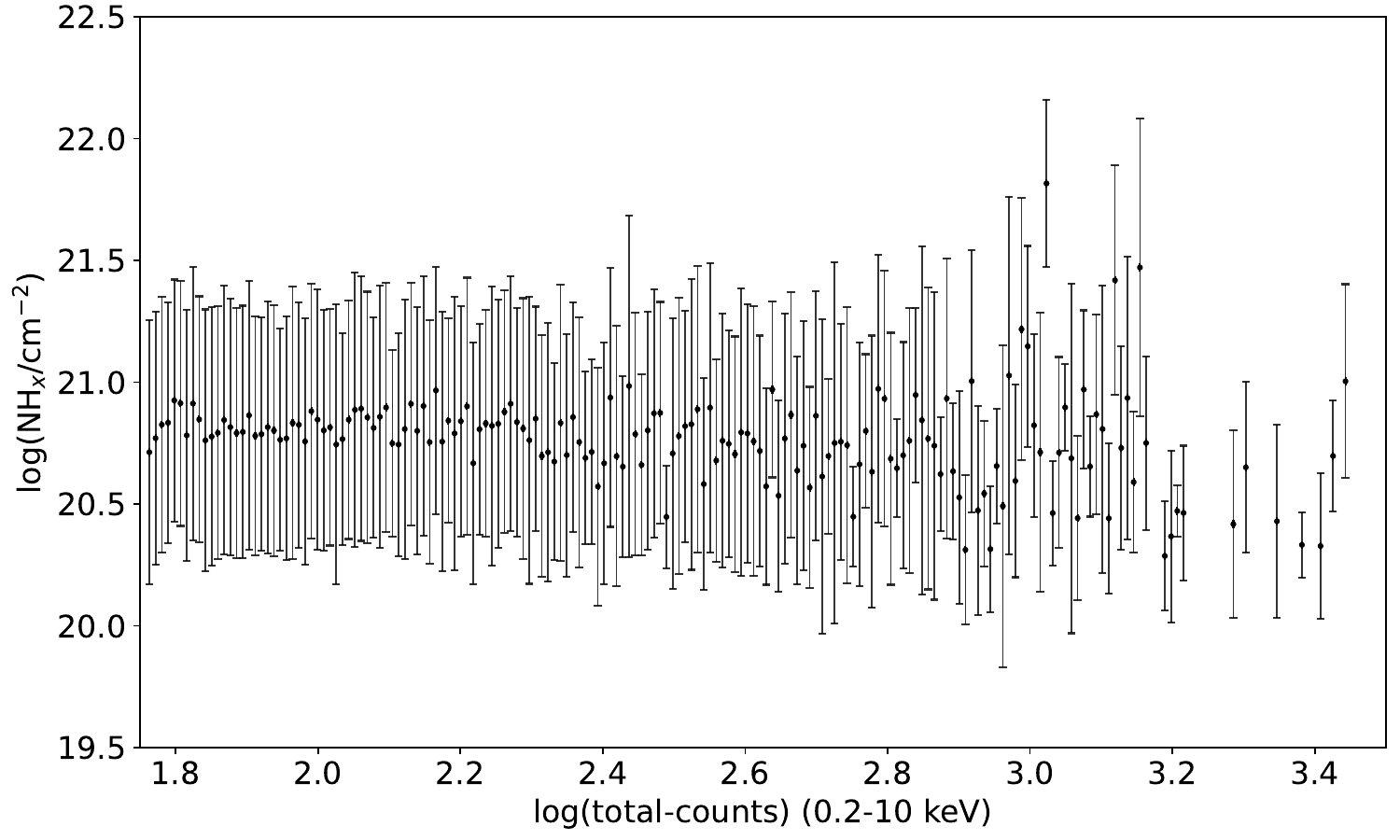}
      \caption{
      NH$_{X}$ distribution as function of the number of counts in the 0.2-10~keV energy range.
 The data has been rebinned for illustrative purposes.      
      }\label{fig_nh_counts}
   \end{figure}

Figure~\ref{fig_nh_counts} shows the NH$_{X}$ distribution as a function of the number of counts in the 0.2-10~keV energy range.
Note that five objects have a number of counts $>\log(3.2)$. 
However, we zoomed in on the plot for clarity.
We have rebinned our dataset in such a way that each bin aggregates approximately 200 data points, for illustrative purposes, with error bars indicating the uncertainties of the sample of sources within each bin. 
The rebinning allows us to focus on the larger-scale features of the data, enabling more straightforward interpretation and analysis.
We note that for this sample, a broad range of NH$_{X}$ measurements can be obtained, and we are able to recover low column densities ($<10^{21}$ cm$^{-2}$).
We found that the average uncertainty for the sample is $35\%$, which decreases to $15\%$ for sources with $>\log(2.5)$ number of counts.
  
\begin{figure}
\includegraphics[width=0.48\textwidth]{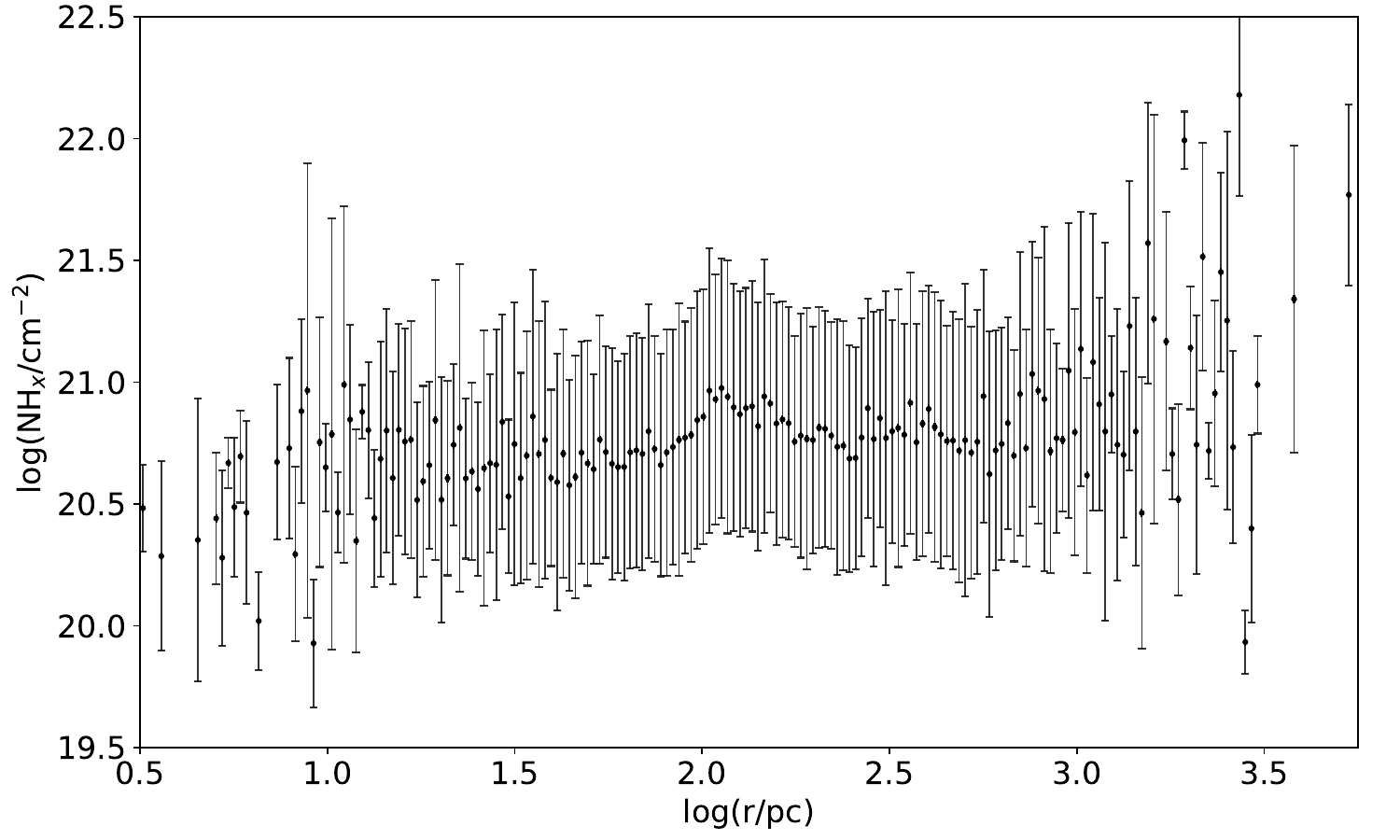}
      \caption{
 NH$_{X}$ as a function of the distances derived from the cross-matching between eRASS1 and {\it Gaia}.
 The data has been rebinned for illustrative purposes.
      }\label{fig_nh_dis}
   \end{figure}

Figure~\ref{fig_nh_dis} shows the NH$_{X}$ distribution as a function of the distance for each source.
The data has been rebinned for illustrative purposes.    
The hydrogen column densities range from  $10^{19}$ -- $10^{22}$ cm$^{-2}$.
There is no clear correlation between the column densities and the distances, which is expected given that not only the distances but also the position of the source (i.e., the celestial coordinates) affects the density of the absorber.

\begin{figure}
\includegraphics[width=0.48\textwidth]{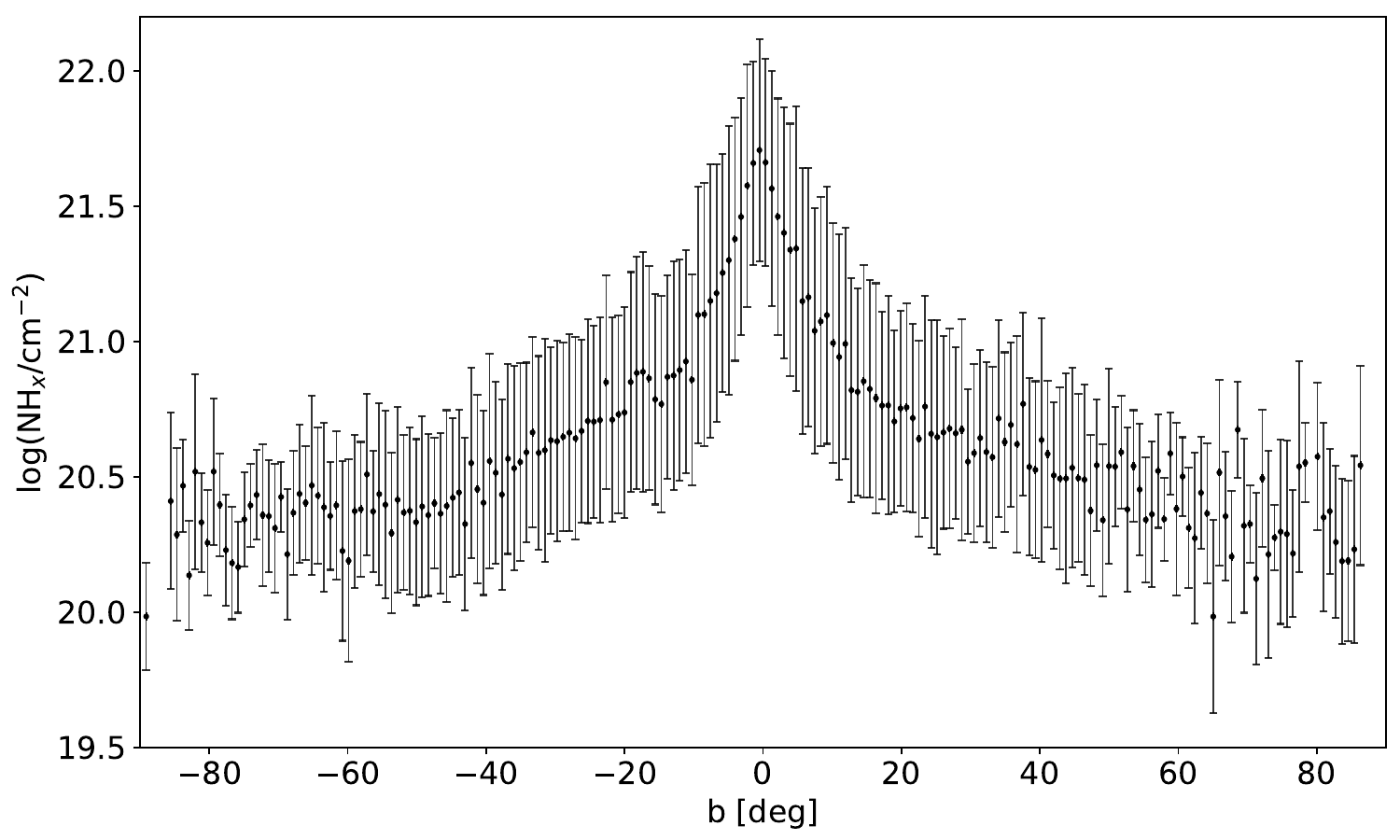}\\
\includegraphics[width=0.48\textwidth]{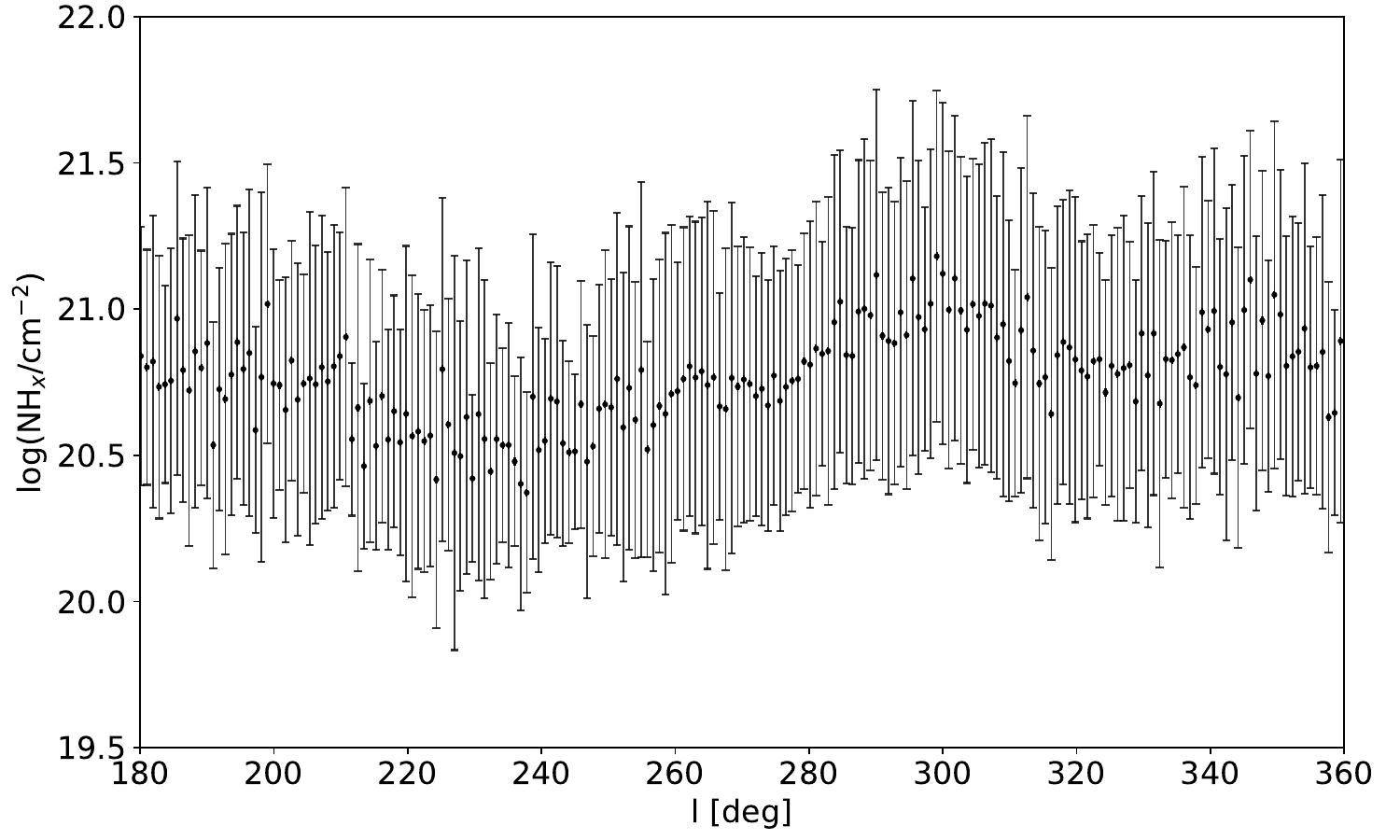}
      \caption{
  \emph{Top panel:} NH$_{X}$ distribution as function of the Galactic latitude.
  \emph{Bottom panel:} NH$_{X}$ distribution as function of the Galactic longitude.
  The data has been rebinned for illustrative purposes. 
      }\label{fig_nh_l_b}
   \end{figure}

Figure~\ref{fig_nh_l_b} shows the column densities distribution as a function of the Galactic latitude (top panel) and Galactic longitude (bottom panel). 
We found a solid correlation of the NH$_{X}$ values with the Galactic latitude, which indicates a decrease of the ISM material density along the vertical direction away from the Galactic plane. 
Establishing a clear relationship between the NH$_{X}$ and the Galactic longitude is difficult.

\subsection{Comparison with previous X-ray ISM absorption analysis}\label{sec_nh_xray}
We have measured a range of column densities for the cold component in good agreement with previous high-resolution spectra analysis \citep[see, for example][]{pin13,gat16,eck17,gat18,gat18b}. 
\citet{gat18}, in particular, found a similar correlation between NH$_{X}$ values and the Galactic latitude/longitude from a sample including both Galactic and extragalactic sources. 
However, most works based on high-resolution spectra used X-ray binaries as Galactic sources, thus avoiding studying regions close to the Galactic plane.
\citet{gat18b} also observed a general trend of higher column densities near the Galactic plane in their analysis.
A primary difference between our results and their measurements is that we can recover low column densities ($<10^{20}$ cm$^{-2}$) for Galactic latitudes (i.e., $|b|>20^{\circ}$).
This is most likely model-related, as \citet{gat18b} only considered single-temperature coronal models, and they fixed the abundance of the Galactic X-ray sources to solar values.

\subsection{Comparison with 21~cm surveys}\label{sec_nh_21cm}
\begin{figure}
\includegraphics[width=0.48\textwidth]{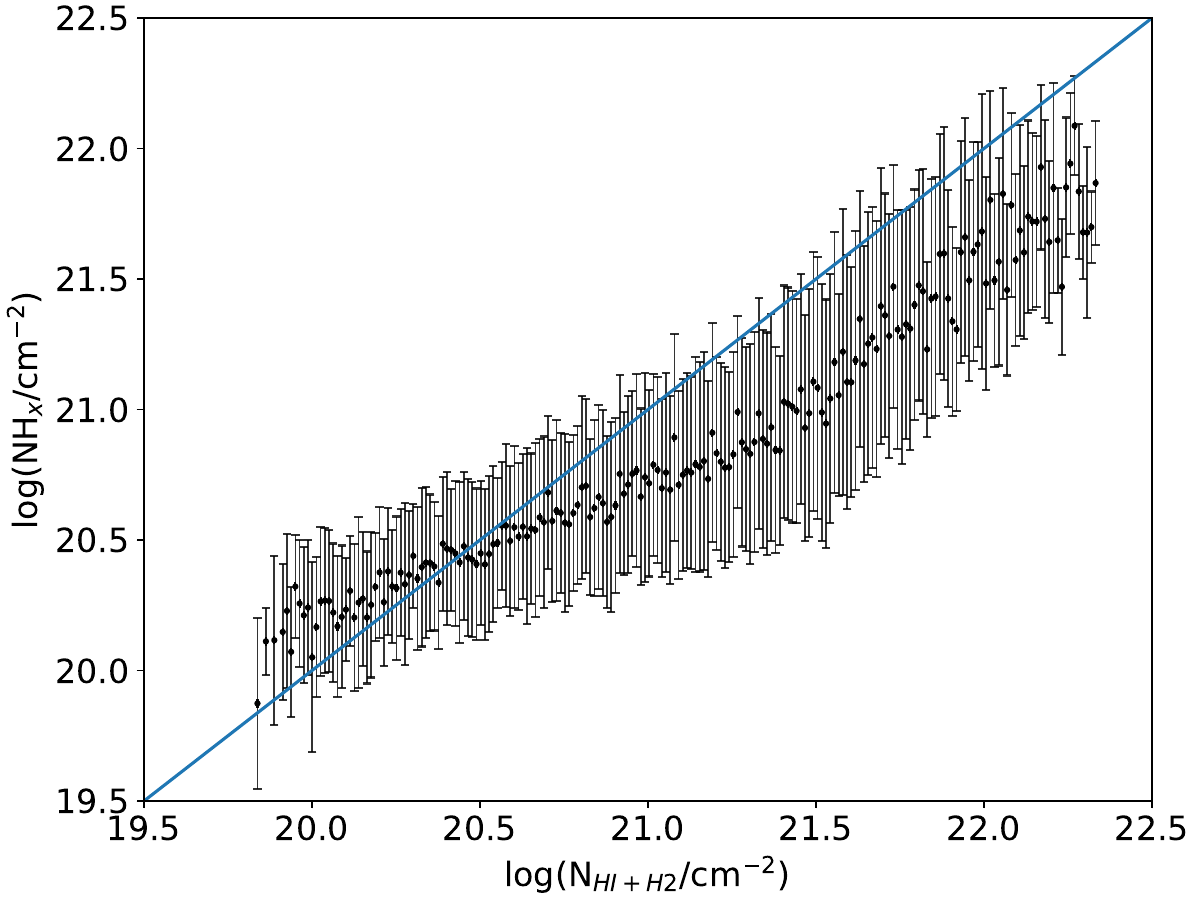}
      \caption{
Comparison between the NH$_{X}$ values obtained from the best-fits and the NH$_{21cm}$ measured in the lines-of-sight.
      The $\log($NH$_{x})$/$\log($NH$_{21cm})$ $=1$ ratio is plotted with a solid blue line.  
The data has been rebinned for illustrative purposes.
      }\label{nh_21cm}
   \end{figure}

Figure~\ref{nh_21cm} shows a comparison between the NH$_{X}$ and the NH$_{21cm}$ measured in the lines-of-sight. 
The data has been rebinned for illustrative purposes.
We obtained the values from \citet{wil13}, which include not only the atomic component but also the molecular component ($H_{2}$).
The $\log($NH$_{x})$/$\log($NH$_{21cm})$ $=1$ ratio is plotted with a solid blue line.  
In general, the values are better distributed around the $\log($NH$_{x})$/$\log($NH$_{21cm})$ unity ratio than \citet[][see their Figure 6]{gat18}.
However, there are high-density points where NH$_{X}$ values tend to be lower than NH$_{21cm}$.
As described above, our NH$_{X}$ measurements allowed us to recover low column densities.
Also, 21~cm maps provide measurements over the entire Galactic line-of-sight. 
Therefore, they can lead to overestimating the column densities for near sources.

Analysis done with high-resolution spectra tends to show lower X-ray column densities than 21~cm measurements \citep[see for example ][]{gat17,gat18}.
Lines-of-sight where NH$_{X}$ is larger than 21~cm measurements may correspond to regions with absorbers other than atomic hydrogen, including partially ionized gas \citep[e.g., ][]{gat16} or molecular absorption \citep{joa16}, a scenario proposed by \citet{wil13}.
Finally, differences between 21~cm and X-ray values may be related to the difference in effective beam sizes.  
While the size of regions probed by X-ray measurements is limited by the size of the distant source, 21~cm measurements come from a beam size of $\sim$0.7 deg$^2$ \citep{kal05}.

\section{The gas-to-extinction ratio}\label{sec_nh_other}
\begin{figure} 
\includegraphics[width=0.48\textwidth]{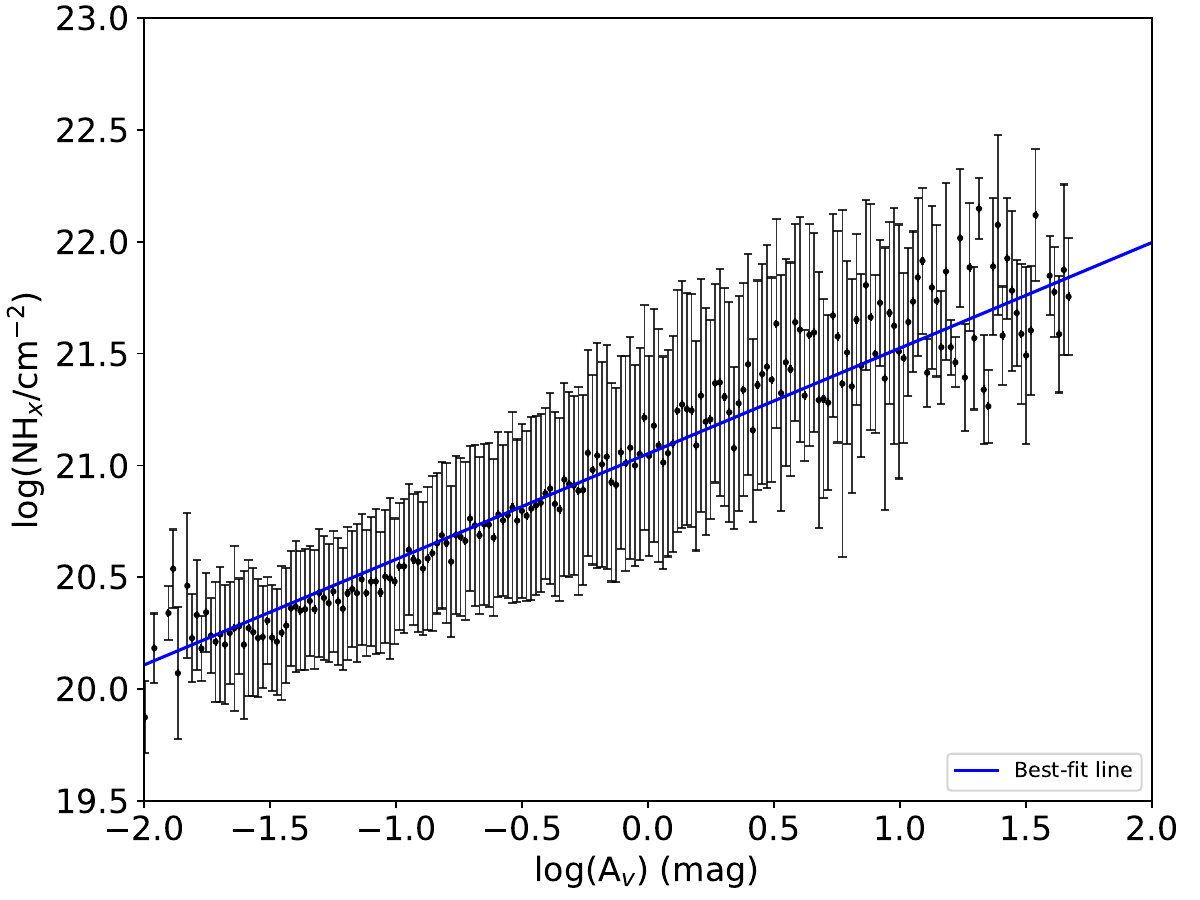}
      \caption{
Distribution of NH$_{X}$ as a function of the extinction in the line-of-sights. 
Black data points correspond to the NH$_{X}$ values obtained from the X-ray spectra.
We used the \citet{sch11} dust survey for the extinction calculation in combination with the SDSS red filter (6231\AA). 
The blue solid line indicates the best-fit linear fit result.
The data has been rebinned for illustrative purposes.
      }\label{fig_nh_dust}
   \end{figure}

Figure~\ref{fig_nh_dust} compares NH$_{X}$ as a function of the extinction in the line-of-sights.
Black data points correspond to the NH$_{X}$ values obtained from the best fit.
Extinction values were computed from \citet{sch98,sch11a} with the SDSS red filter (6231\AA).
We computed a linear fit to the data in the form:
\begin{equation}
NH_{X}=MA_{V} + B
\end{equation}
The best-fit parameters are $M=0.46\pm 0.02$ (mag) and $B=0.09\pm 0.01$ ($\times 10^{21}$ cm$^{-2}$). 
The blue solid line shows the best-fit result.
The slope found from the best fit is larger than the value obtained by \citet[][$M=0.225\pm 0.003$]{zhu17}. 
However, their sample is much smaller ($\sim 100$ sources) and includes supernova remnants (SNRs), X-ray binaries (XBs), and planetary nebulae (PNe). 
Therefore, intrinsic absorption may affect their measurement.
Also, their sources cover Galactic latitude ranges of $|b|<20^{\circ}$, thus limiting their analysis to the Galactic plane, but covering distances up to $\sim13$~kpc.
On the other hand, our best-fit parameter is in good agreement with the {\it ROSAT} measurements obtained by \citet{pre95}.
Our linear fit appears to underestimate the data quite strongly at larger $A_{V}$.
Some potential reasons include saturation effects from the X-ray absorption, variability in dust properties (due to composition, size distribution etc.) across different lines of sight, non-linear effects or variations due to the complex interplay between gas and dust phases and physical processes such as clumpiness or variations in metallicity.

\begin{figure} 
\includegraphics[width=0.48\textwidth]{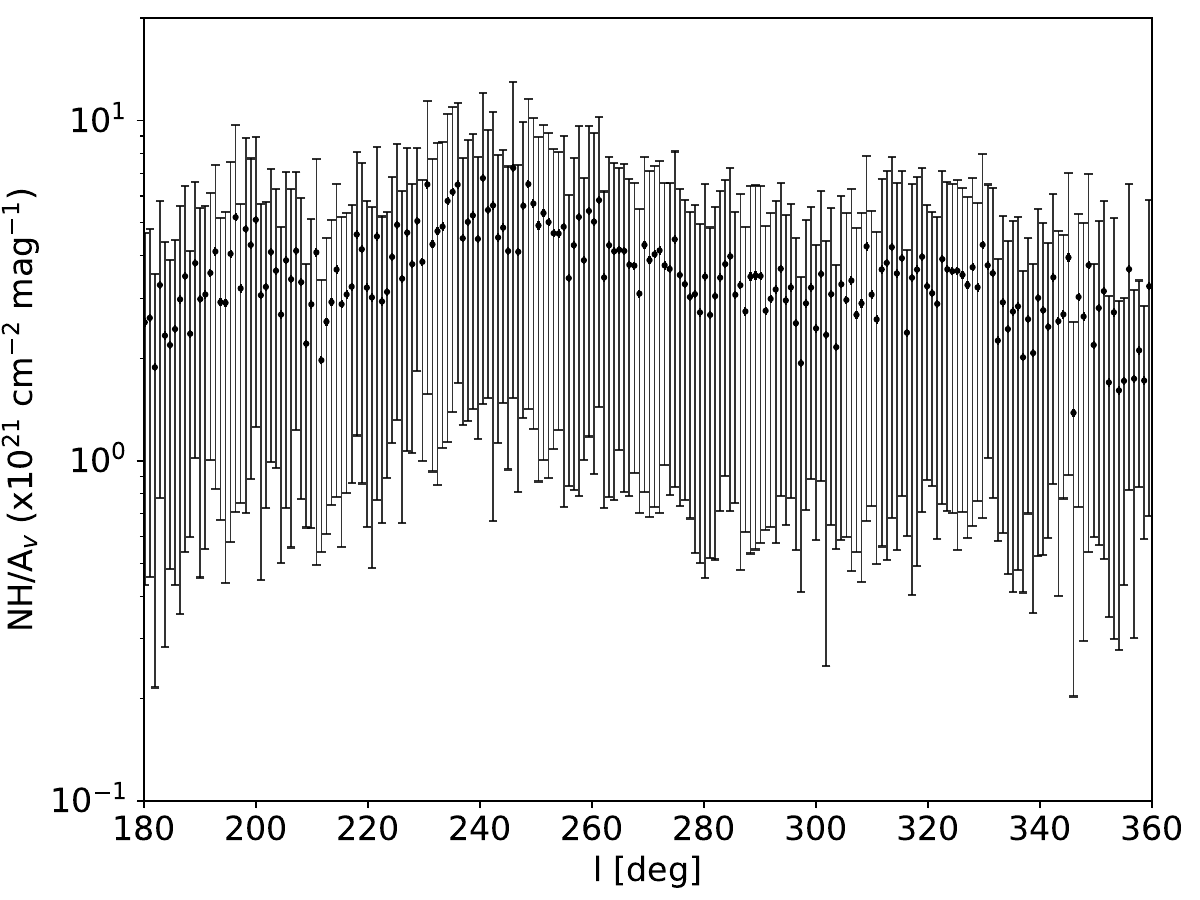}\\
\includegraphics[width=0.48\textwidth]{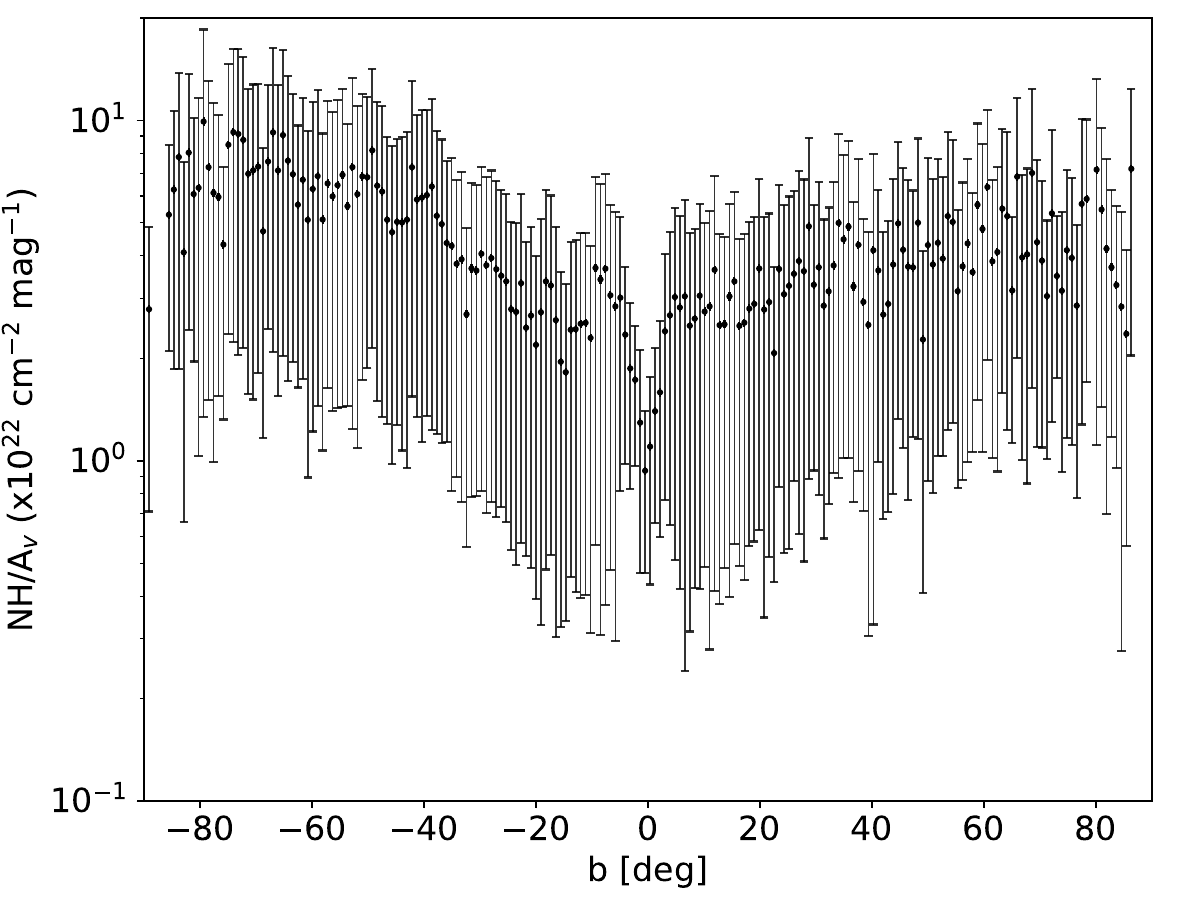}
      \caption{
Distribution of the NH$_{X}$/A$_{V}$ ratio as a function of the Galactic longitude (top panel) and Galactic latitude (bottom panel) in the line-of-sights. 
The data has been rebinned for illustrative purposes.
      }\label{fig_dust_lat_lon}
   \end{figure}

Figure~\ref{fig_dust_lat_lon} shows the distribution of the NH$_{X}$/A$_{V}$ ratio as a function of the Galactic longitude (top panel) and Galactic latitude (bottom panel) in the line-of-sights. 
While there is no clear indication for variations along the Galactic longitude, the NH$_{X}$/A$_{V}$ value decreases as we move towards the Galactic plane. 
This may reflect the increase of the dust ISM component towards the Galactic plane. \citep{dam01,wei01,gre18}.

\begin{figure} 
\includegraphics[width=0.48\textwidth]{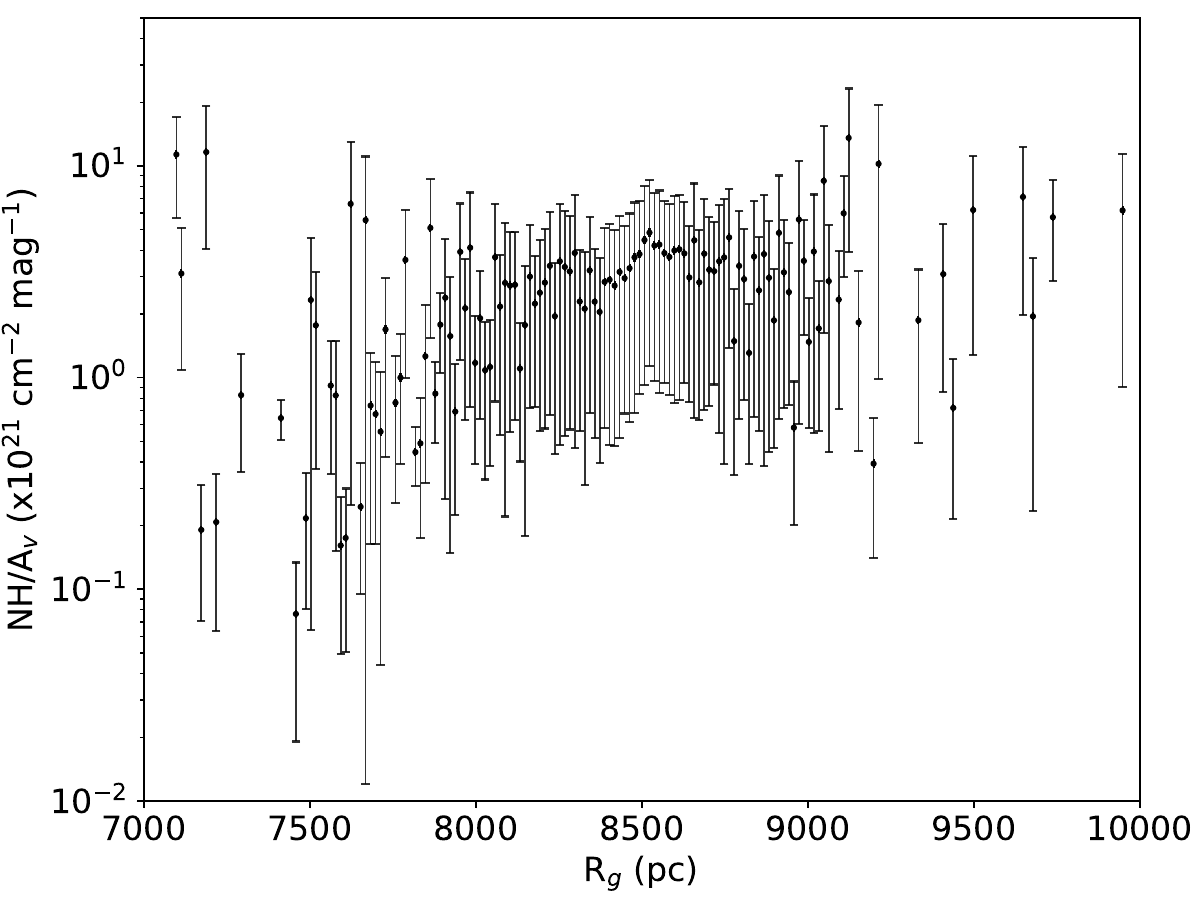}\\
\includegraphics[width=0.48\textwidth]{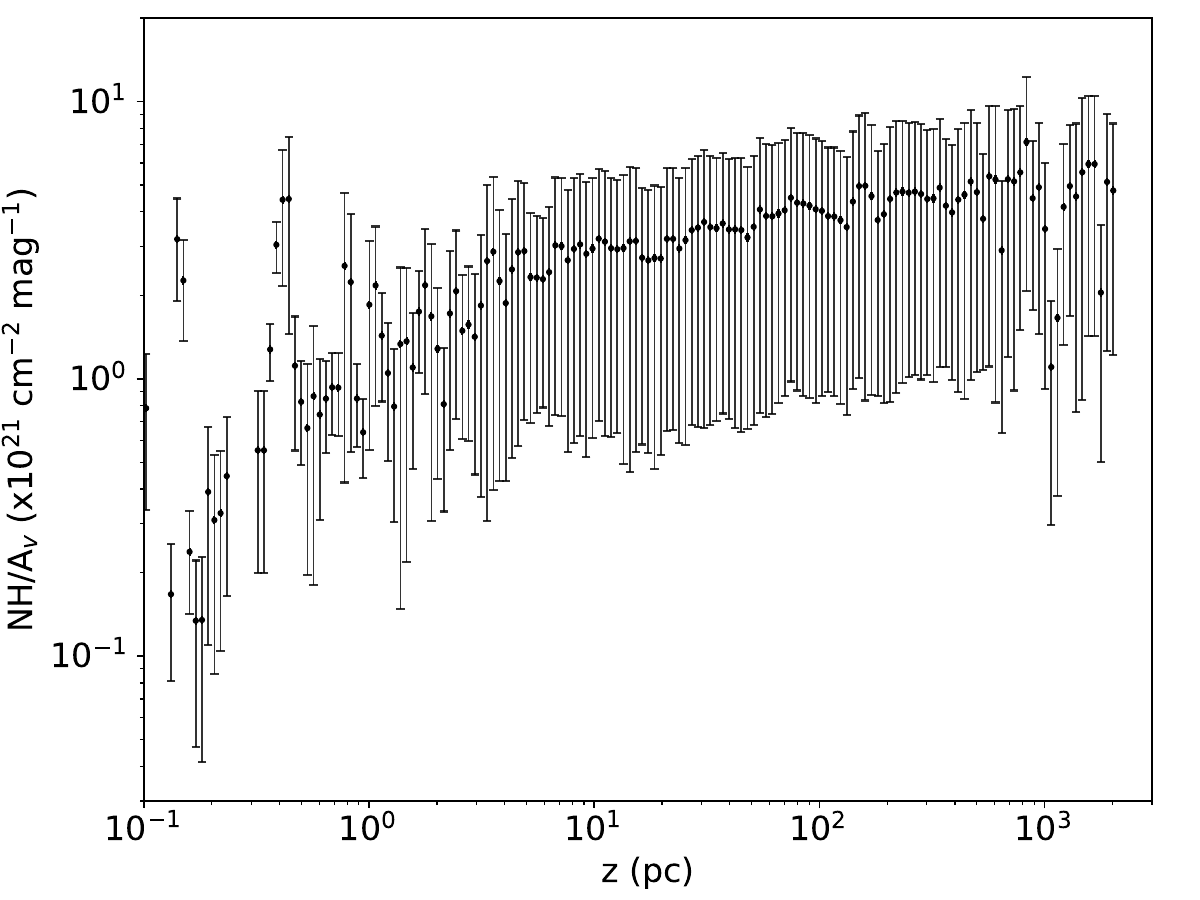}\\ 
\includegraphics[width=0.48\textwidth]{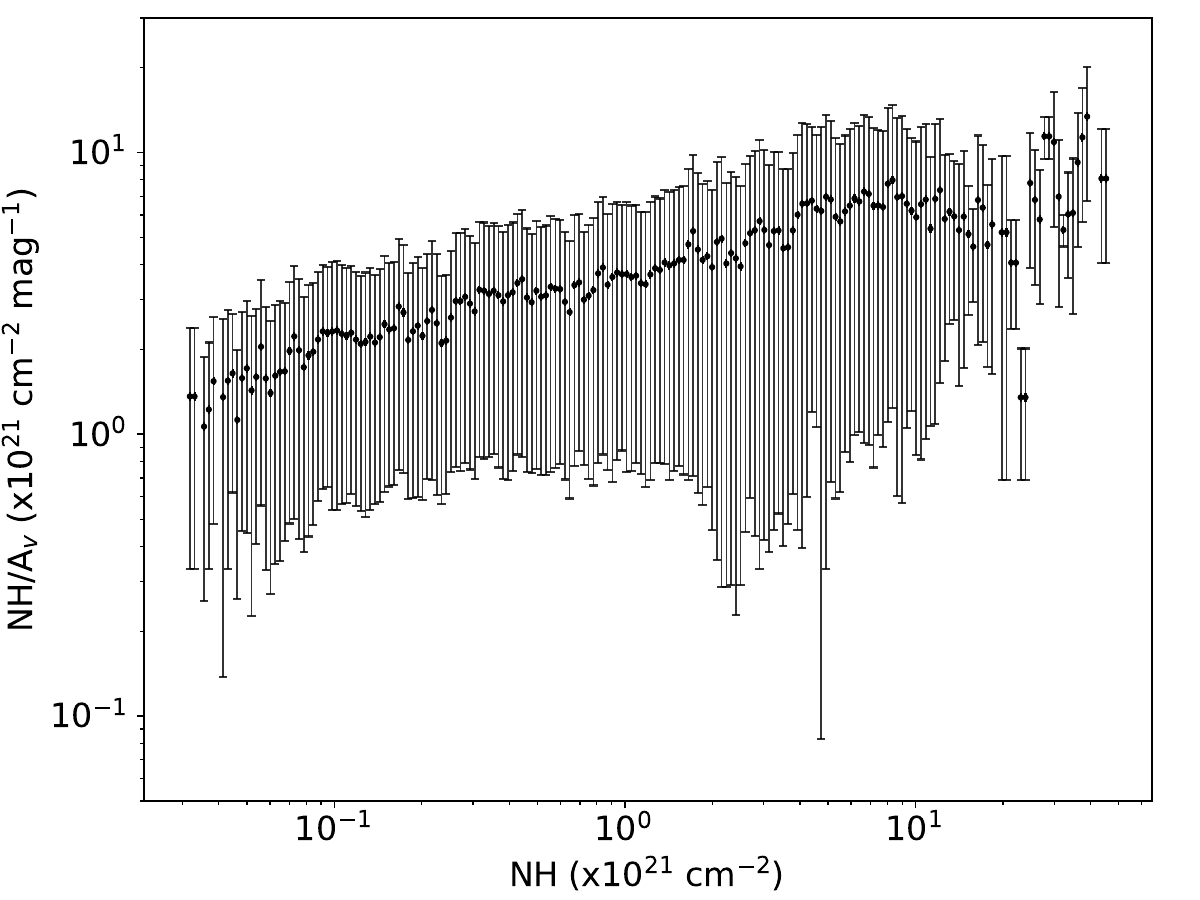}
      \caption{
Distribution of the NH$_{X}$/A$_{V}$ ratio as a function of $R_{g}$ (top panel), $z$ (middle panel), and NH$_{X}$ (bottom panel). 
The data has been rebinned for illustrative purposes.
      }\label{fig_dust_geom}
   \end{figure}

Figure~\ref{fig_dust_geom} shows the NH$_{X}$/A$_{V}$ ratio as function of (a) the distance from the Galactic Center, assuming $R_{sun}=8.5$~kpc ($R_{g}$); (b) the distance from the Galactic plane ($z$) and (c) NH$_{X}$.
There is a hint for an increasing NH$_{X}$/A$_{V}$ as a function of $z$ and NH$_{X}$. However, the uncertainties are significant.

\section{Neutral absorption density laws}\label{sec_law}
\begin{figure}
\includegraphics[width=0.48\textwidth]{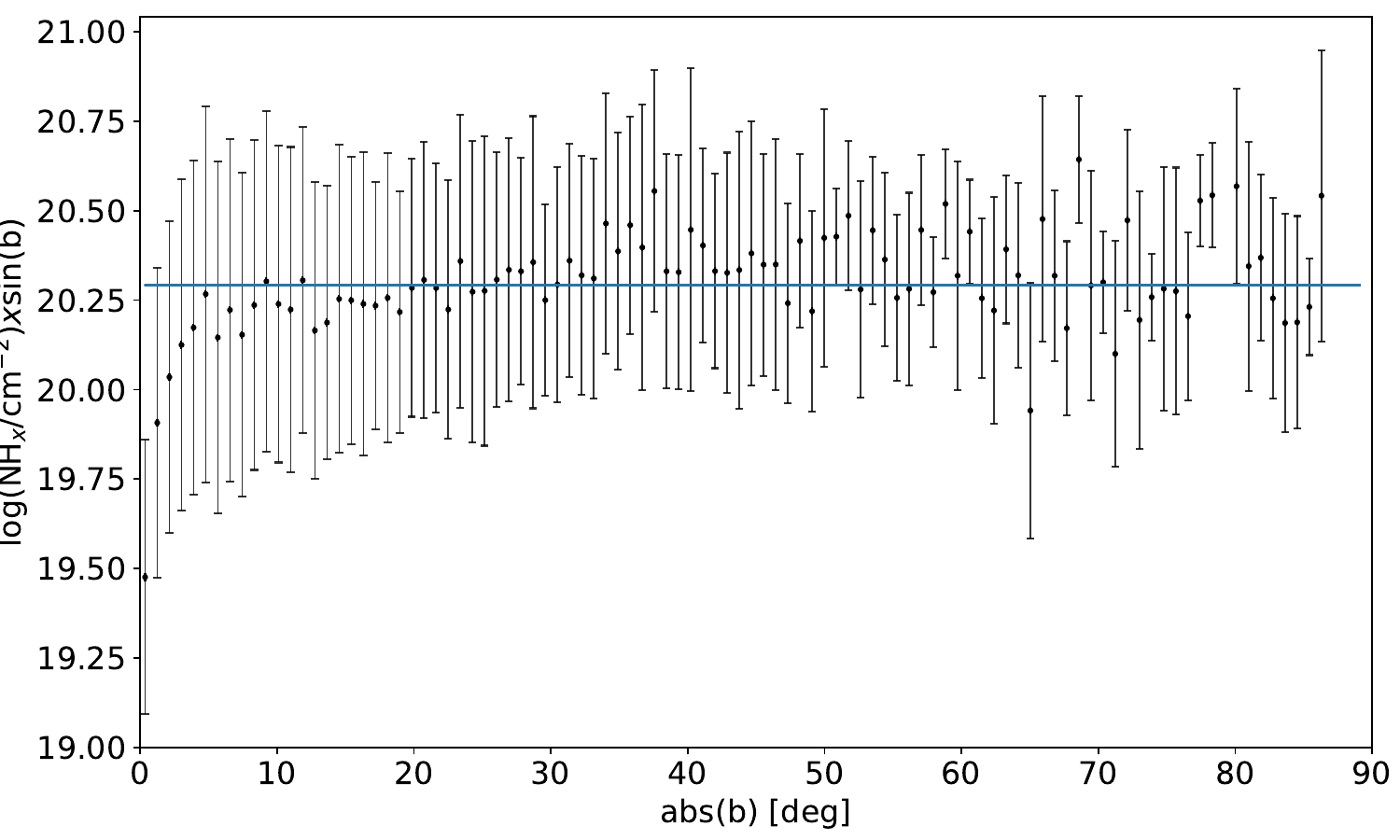} 
      \caption{X-ray column densities multiplied by the $|sin(b)|$ as a function of $|b|$. 
      The blue horizontal line indicates the mean value.
      In a infinitely thin disc model such product should be the same for all sources.
      }\label{fig_nh_sinb_b}
   \end{figure}

Using the equivalent column densities derived from the X-ray fits, the neutral gas distribution can be modeled according to the equation
\begin{equation}
NH_{X}(\vec{r}_{source}) = \int_{\vec{r}_{observer}}^{\vec{r}_{source}} n(\vec{r}) d\vec{r}
\end{equation} 
where $r_{source}$ is the distance along the line of sight to a given source and $n(r)$ is the density law of the neutral gas in the Milky Way. 
First, we assumed the simplest possible model, which consists of a very thin disc for which column densities will follow a simple law NH$_{X}=$N$_{Z}/|\sin(b)|$, where $b$ is the Galactic latitude and N$_{Z}$ is the vertically integrated disc column density. 
Figure~\ref{fig_nh_sinb_b} shows the values of N$_{Z}=$NH$_{X}|\sin(b)|$ as function of $|b|$.
We found a mean value of N$_{Z}=1.93\times 10^{20}$ cm$^{-2}$ (indicated by the blue horizontal line).
Although we have identified a strong dependence of NH$_{X}$ on $b$ (Figure~\ref{fig_nh_l_b}), the plot shows small differences in N$_{Z}$, thus indicating that an infinitely thin disc
model can be appropriated.
This is expected because we are modeling column densities within the Galactic plane but at large distances from the Galactic center.
Indeed, the minimum distance between the sources in our sample and the center of the Milky Way is $\sim5380$~pc.

We test different density models to determine whether we can obtain constraints in the density law parameters.
Given that most of the ISM radial profiles are better described in cylindrical coordinates, we define the Galactocentric coordinates $\left(R,z\right)$ as:
\begin{align}
R^{2}=r^{2}\cos^{2}(b)-2rR_{\odot}\cos(b)\cos(l)+R_{\odot}^{2}\\
z=r\sin(b)
\end{align}
where $R_{\odot}$ is the distance between the Sun and the Galactic center. 
We have tested a family of density profiles commonly used to model the cold gas contribution in the Milky Way. 
\begin{enumerate}
\item Model A: as described by \citet{rob03}
\begin{align}
&&n_{A}(R,z)=n_{0}\times \exp{\left(\frac{R}{h_{R}}\right)}\times\exp{\left(-\frac{|z|}{h_{z}}\right)}
\end{align}
where the central density ($n_{0}$), the core radius ($h_{R}$) and the core height ($h_{z}$) are the free parameters of the model.

\item  Model B: as described by \citet{mar10}
\begin{align}
&&n_{B}(R,z)=n_{0}\left(1+\frac{R}{R_{g}}\right)^{\gamma}\times\exp^{-R/R_{g}}\times \sech^{2}\left(\frac{z}{h_{g}}\right)
\end{align}
where the scaleheight $h_{g}$ depends on $R$ as:
\begin{align}\label{equ_hg}
&&h_{g}(R)=h_{0}+\left(\frac{R}{h_{R}}\right)^{\delta}
\end{align}
The free parameters of the model are the central density ($n_{0}$), the core radius ($R_{g}$), the slope ($\gamma$), and the scaleheight parameters ($h_{R},h_{R}$ and $h_{0}$).

\item  Model C: as described by \citet{mar13}
\begin{align}
&&n_{C}(R,z)=n_{0}\left(1+\frac{R}{R_{g}}\right)^{\gamma}\times\exp^{-R/R_{g}}\times \exp^{-z/h_{g}}
\end{align}
where the scaleheight $h_{g}$ is defined as Equation~\ref{equ_hg}. 
The free parameters of the model are the central density ($n_{0}$), the core radius ($R_{g}$), the slope ($\gamma$), and the scaleheight parameters  ($h_{R},h_{R}$ and $h_{0}$).

\item Model D: as described by \citet{bar16}
\begin{align}
&&n_{D}(R,z)=\frac{n(R)}{2.12z_{0}}\times\exp{\left[-\left(\frac{z}{1.18z_{0}}\right)^{2}\right]}
\end{align}
with $n(R)$ described as
\begin{align}
&\rho(R)=n_{0}\times  \exp{\left[-\frac{\left(R^{3/2}-R_{0}^{3/2}\right)}{R_{0'}^{3/2}}-R_{0''}\left(\frac{1}{R^{2}}-\frac{1}{R_{0}^{2}}\right)\right]}
\end{align}
The free parameters of the model are the central density ($n_{0}$), the core radius parameters ($R_{0},R_{0'},R_{0''}$) and the core height ($z_{0}$).

\item Model E: as described by \citet{nic16}
\begin{align}
&&n_{E}(R,z)=n_{0}\times\exp{\left(-\sqrt{(R/h_{R})^{2}+(z/h_{z})^{2}}\right)}
\end{align}
The free parameters of the model are the central density ($n_{0}$), the core radius ($R_{c}$) and the core height ($h_{z}$).

\item Model F: as described by \citet{gat18}
\begin{align}
&&n_{F}(R,z)=n_{0}\times\exp{\left(-R/h_{R}\right)}+\exp{\left(-|z|/h_{z}\right)}
\end{align}
The free parameters of the model are the central density ($n_{0}$), the core radius ($R_{c}$) and the core height ($h_{z}$).
\end{enumerate}

\begin{figure*}
\includegraphics[width=0.98\textwidth]{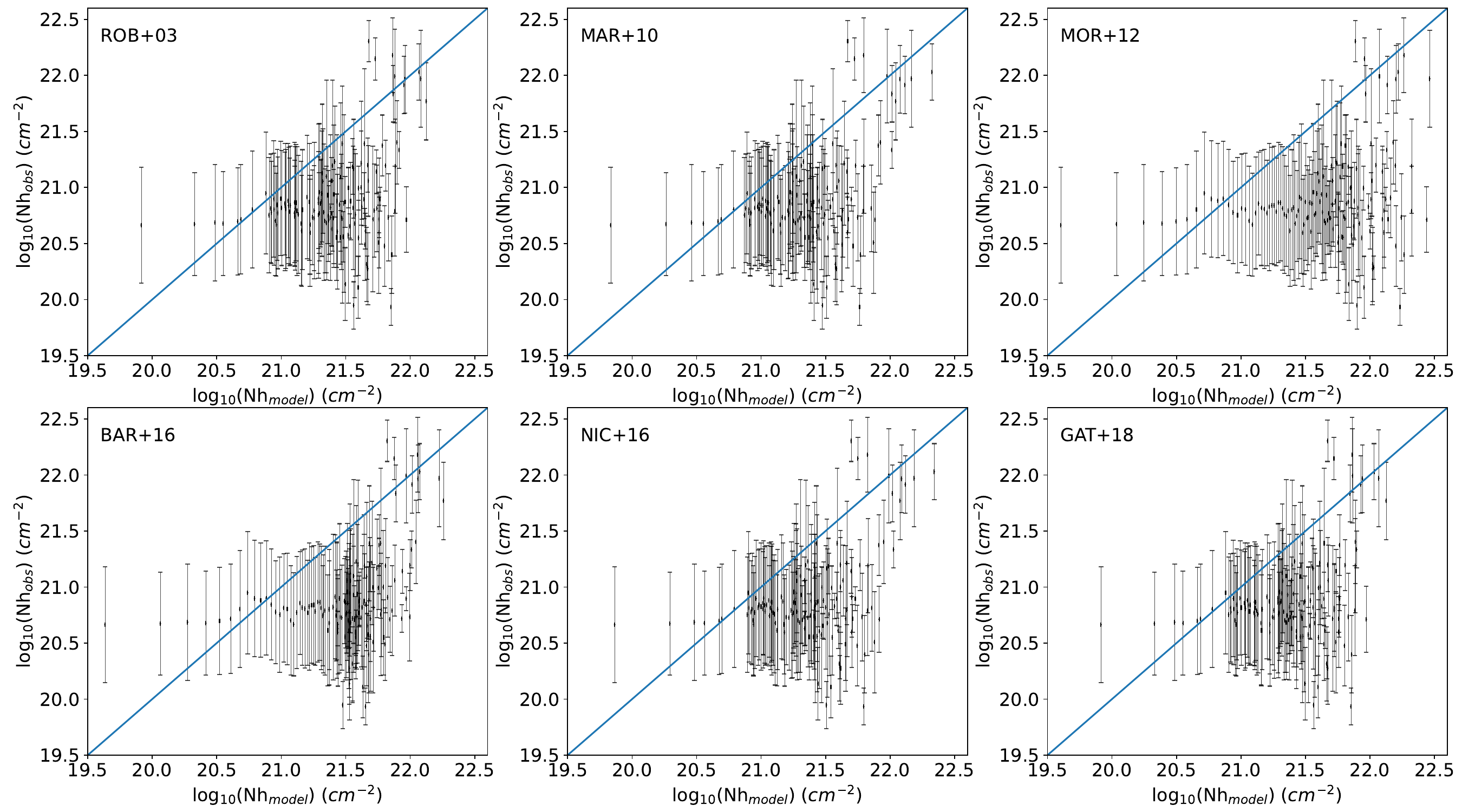} 
      \caption{Comparison between the measured and the model-predicted column densities for the different density laws described in Section~\ref{sec_law}. It is clear that none of the model can reproduce the full complexity of the measurements. }\label{fig_nh_law_fit}
   \end{figure*}

For each model the parameters change and then a Pearson's chi-squared test was calculated according to the formula
\begin{align}
&&\chi^{2}=\sum_{i=1}^{N}\frac{(NH_{{\rm X},i}-NH_{{\rm n},i})^{2}}{\sigma_{{\rm X},i}^{2}}
\end{align}
where $NH_{{\rm X},i}$ is the measured column density from the X-ray observations, $NH_{{\rm n},i}$ is the expected column density value from the analytical profiles, $\sigma_{{\rm X},i}$ is the column density uncertainty and $1 \le i \le N $ indicates the source. 
Table~\ref{tab_density} shows the best-fit parameters and the $\chi^{2}$ value obtained for each density profile. 
Figure~\ref{fig_nh_law_fit} compares the predicted and measured column densities. 
As mentioned before, it is important to note that we are modeling the local ISM around the Sun, and our sample lacks sources near the Galactic center. 
Also, the density profiles predict a smooth distribution of the column densities over the sky. At the same time, 21 cm observations indicate substantial intensity variations on different angular scales.
Therefore, we can expect relatively weak constraints in the radial scale ($R_{c}$) and on the central density (as seen by the dispersion of these parameters between the models).
All best-fits lead to poor best-fit statistics, although Model D performs slightly better.
However, we noted that the thickness of the disc ($h_{z}$) tends to be well constrained among different models with values in the $8 < h_{z} < 30$~pc range, in good agreement with the values obtained by \citet{gat18} in their analysis of the X-ray ISM absorption using Galactic and Extragalactic sources.
We also tested fitting the data by fixing the radial scale parameters, and we have found similar values for the other parameters.

\begin{table*}
\caption{\label{tab_density}Density laws best-fit parameters.}
\small
\centering
\begin{tabular}{lccccccc}
\hline
Parameter  &Units  & Model A             & Model B        & Model C     & Model D & Model E & Model F\\
\hline
\multicolumn{3}{c}{Model A }\\
$n_{0}$&cm$^{-3}$  &  $0.0038\pm 0.0018$ & $2.32\pm 1.98$ &  $0.053\pm 0.025$ &  $2.54\pm 0.74$&  $5.15\pm 1.11$&  $(2.03\pm 0.84)\pm 10^{-4}$\\ 
$h_{R}$ &pc        &  $1370\pm 109$      & $<9372$        &  $10\pm 3$ &                &  $<7981$&  $ 929\pm 42$\\ 
$h_{z}$ &pc        &  $18\pm 2$          &                &        &                &  $11\pm 3$&  $28 \pm 2$\\ 
$h_{0}$ &pc        &                     & $<3281 $       &  $<2228$ &                &       &       \\ 
$\gamma$ &         &                     & $4.97\pm 0.30$ &  $4.98\pm 0.35$ &                &       &       \\ 
$\delta$ &         &                     & $<10$          &  $<10$ &                &       &       \\ 
$R_{g}$&pc         &                     & $556\pm 2$     &  $<6646$ &                &       &       \\  
$R_{0}$ &pc        &                     &                &        &  $9262\pm 607$ &       &       \\ 
$R_{0'}$ &pc       &                     &                &        &  $5496\pm 391$ &       &       \\ 
$R_{0''}$ &pc      &                     &                &        &  $<10000$      &       &       \\ 
$z_{0}$ &pc        &                     &                &        &  $16\pm 1$     &       &       \\ 
$\chi^{2}$/d.o.f&  & 46132/8228          &  41770/8225    &  42942/8225     &  35130/8226     &   46476/8228   &   46108/8228  \\  
\hline    
\multicolumn{3}{l}{d.o.f = degrees of freedom}
\end{tabular}
\end{table*}

\section{3D mapping of the ISM neutral absorption}\label{sec_map}
To construct a three-dimensional density map of neutral absorption in the local ISM, we adopt the methodology outlined by \citet{dhar22}. 
This approach employs a Gaussian process (GP) to predict hydrogen densities based on hydrogen column densities observed along multiple lines of sight.
The process involves obtaining measurements and uncertainties of absorption towards a sample of stellar sources, along with their three-dimensional positions. 
Subsequently, the logarithm of the density ($\log_{10}(\rho)$) is modeled to ensure positivity and incorporate a GP prior, accounting for correlations among densities in three dimensions. 
The posterior in $\log_{10}(\rho)$ is approximated as a normal distribution, with the prior on $\log_{10}(\rho)$ modeled using a GP with a constant mean.

As described in \citet{dhar22}, the GP hyperparameters include three physical scale lengths in Heliocentric Cartesian coordinates ($x, y, z$) of the physical space, the mean density, and an exponential scale factor. 
These hyper-parameters generate the GP prior, from which sets of priors are predicted. 
The GP is evaluated on a grid in Galactic coordinates ($l, b, d$) to facilitate line-of-sight integration. 
The coordinates of the grid cell centers are then transformed into $x, y, z$ coordinates. 
To facilitate comparison with observed hydrogen column densities, the density is integrated along the line of sight by exponentiating the distribution of priors to obtain $\rho$. 
Subsequently, numerical integration along the line-of-sight to all observed coronal sources yields integrated densities, which are compared to observed column densities through likelihood optimization of the model. 
This hierarchical approach treats the GP as a prior on (the logarithm of) hydrogen column density, whose plausible values are determined by the hyper-parameters. 
The optimization process refines hydrogen density as a function of position. 
The algorithm is implemented using the {\tt GPyTorch} package \citep{gard21} and the probabilistic programming package {\tt Pyro} \citep{pha19}, both built upon the {\tt PyTorch} \citep{pas19} machine learning framework. 
The source code has been adapted to directly work with hydrogen column densities instead of dust extinction density.

We divided the sample into different regions for which the GP was applied to optimize the calculation. 
Specifically, we divided the physical space into six layers based on distances, as illustrated in Figure~\ref{fig_nh_ait}. 
The model setup also encompassed crucial parameters, including the learning rate ($0.01$), number of iterations ($1000$), number of inducing points ($1000$), and the number of cells for each region along the $x$, $y$, and $z$ directions ($110,110,110$).
The boundaries for each region, along with the scale length, mean density, and scale factor derived from the GP analysis, are detailed in Table~\ref{tab_gas}.
The GP hyper-parameters obtained are within the range found by \citet{dhar22}, in their analysis of Galactic molecular clouds and complex.
Figure~\ref{fig_3dmap1} presents the hydrogen density map sampled at varying distances, akin to a computerized axial tomography. 
Multiple beams and clouds of diverse sizes are evident along all lines of sight, indicative of small-scale structures. 
The color scale represents densities ranging from $n=0$ cm$^{-3}$ to approximately $90$ cm$^{-3}$.

A substantial horizontal cloud with high density is discernible at around $100$~pc, which is also observed in Figure~\ref{fig_nh_dis}, corresponding to the region with the highest source density in our sample (refer to Figure~\ref{fig_his}). Notably, a similar feature has been identified in previous maps of dust reddening and is potentially associated with the Perseus arm of the Milky Way \citep[e.g.,][]{gre19}. 
For distances beyond $1200$~pc, the reduced number of sources results in a density map lacking distinct structures, becoming nearly homogeneous. 
Extrapolating the density profile distribution without a reliable estimate is unfeasible in such scenarios.
Other noteworthy features include high-density regions located at $180\leq l \leq 217$, $-25.5\leq b \leq -3.8$, and $250$~pc $\leq d \leq 550$~pc, potentially associated with the Orion star-forming region. 
Additionally, at $290\leq l \leq 308$, $-22\leq b \leq -10$, and $50$~pc $\leq d \leq 300$~pc, high-density regions may be associated with the Chamaeleon complex. 
However, due to the limited number of sources in these specific regions, a direct comparison with the results obtained by \citet{dhar22} cannot be performed.
Observational constraints, as demonstrated in our analysis, are vital for comparisons with high-resolution 3D hydrodynamical simulations conducted in recent decades \citep[e.g.,][]{dea00, dea01, gen13, lag13, hir16}.

\begin{table*}
\caption{\label{tab_gas}Summary of Gaussian process parameters.}
\small
\centering
\begin{tabular}{lccccccccccccccccccccccc}
\hline
Parameter  &Region 1&Region 2&Region 3&Region 4&Region 5&Region 6 \\
\hline  
$d$ bounds  & 0$\leq d \leq$ 100 & 100$\leq d \leq$ 250& 250$\leq d \leq$ 500 & 500$\leq d \leq$ 750 & 750$\leq d \leq$ 1250 & 1250$\leq d \leq$ 3000 \\ 
$\log_{10}$(Mean density)   & -2.65     & -3.1       & -3.1       & -3.5       & -3.65      & -3.4\\
$\log_{10}$(Scale factor)         & -1.25     & -1.35      & -2.0       & -1.65      & -1.8       & -2.1\\
Scale length ($x,y,z$)     & (20,22,15)& (15,14,14) & (23,23,24) & (75,77,60) & (60,68,64) & (34,35,33)\\ 
\hline    
\multicolumn{7}{l}{$d$ bounds in units of pc. Mean density in units of ($cm^{-3}$). Scale length in units of pc.}
\end{tabular}
\end{table*}

   \begin{figure*}
\includegraphics[width=0.98\textwidth]{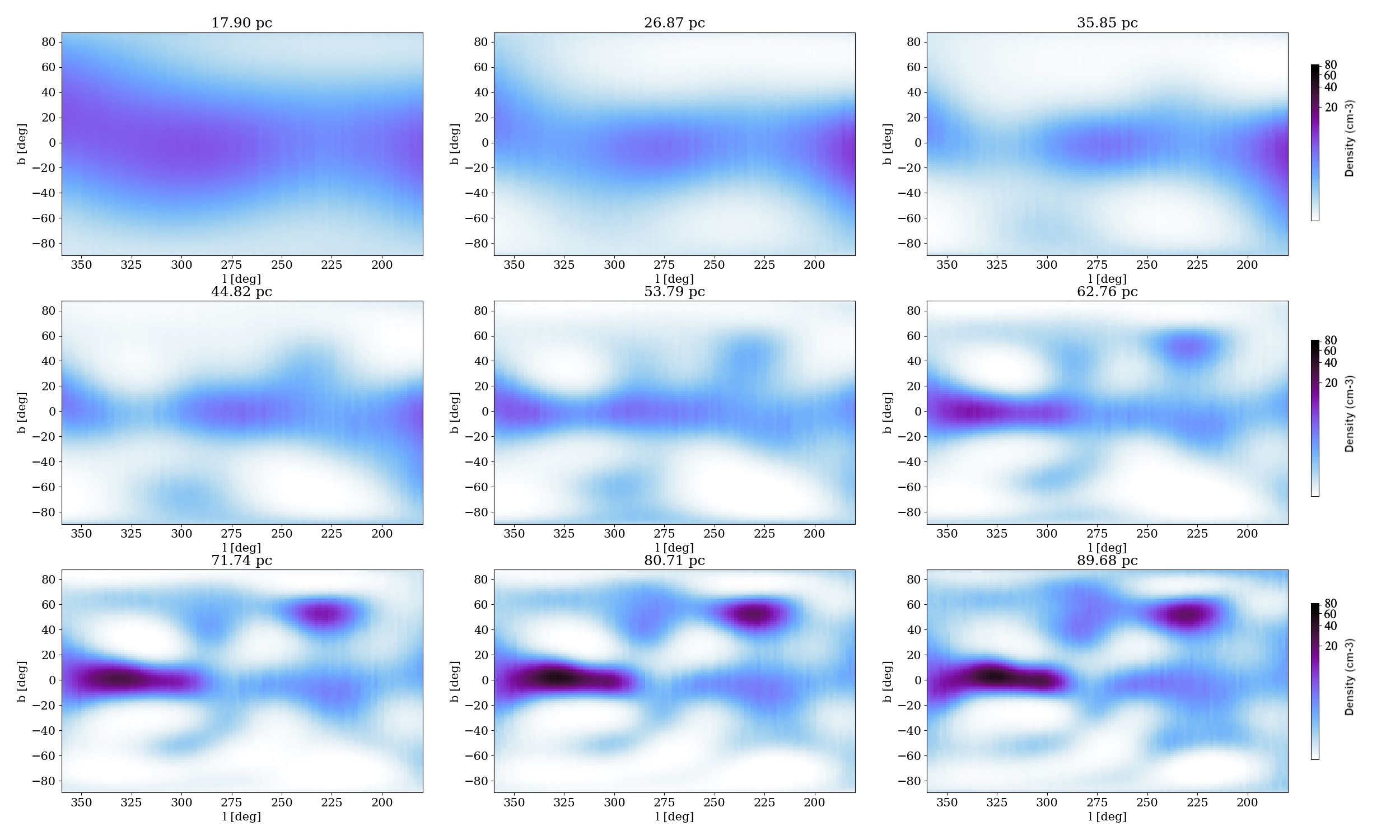} \\
\includegraphics[width=0.98\textwidth]{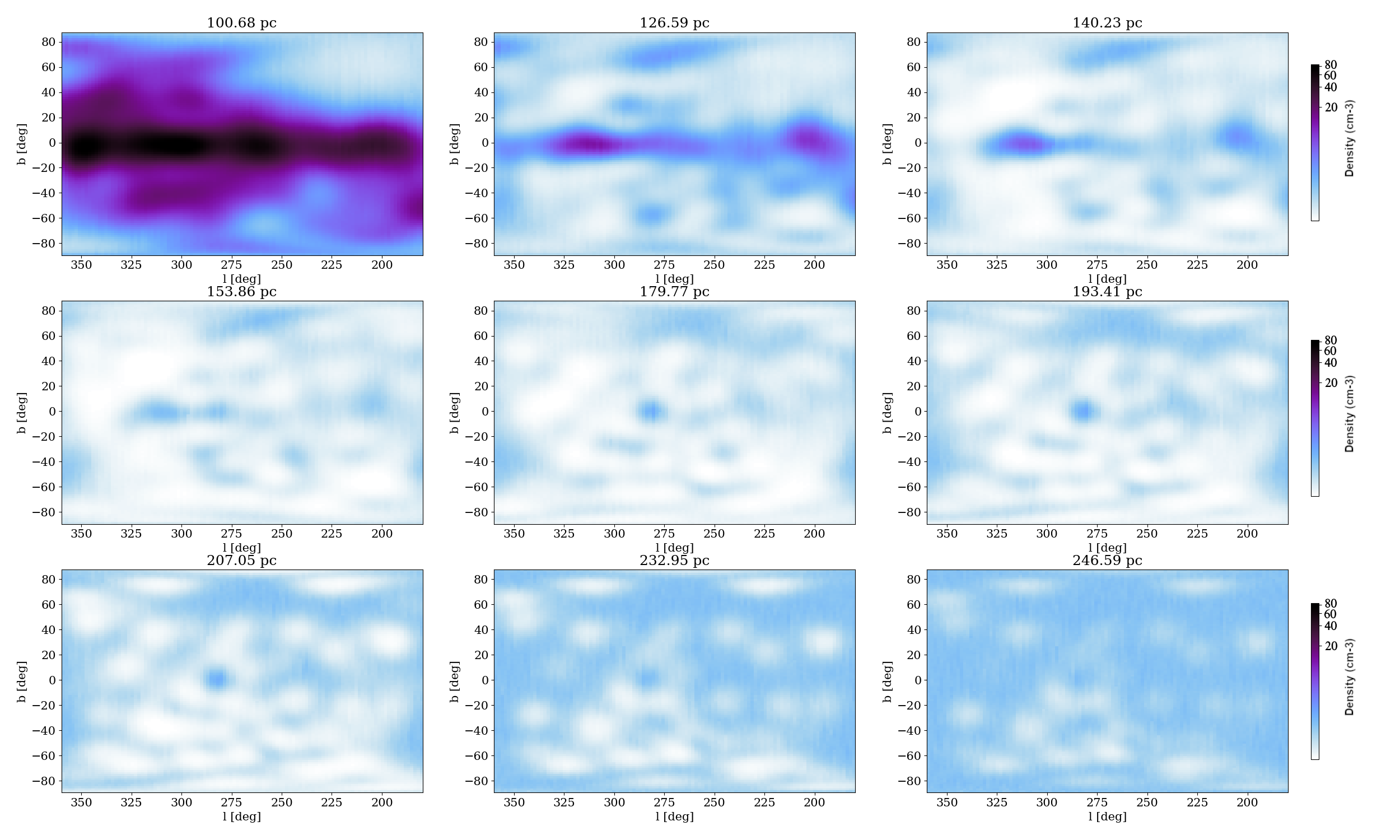} 
      \caption{Hydrogen density map sampled at the indicated distances.
      }\label{fig_3dmap1}
   \end{figure*}

   \begin{figure*}\ContinuedFloat 
\includegraphics[width=0.98\textwidth]{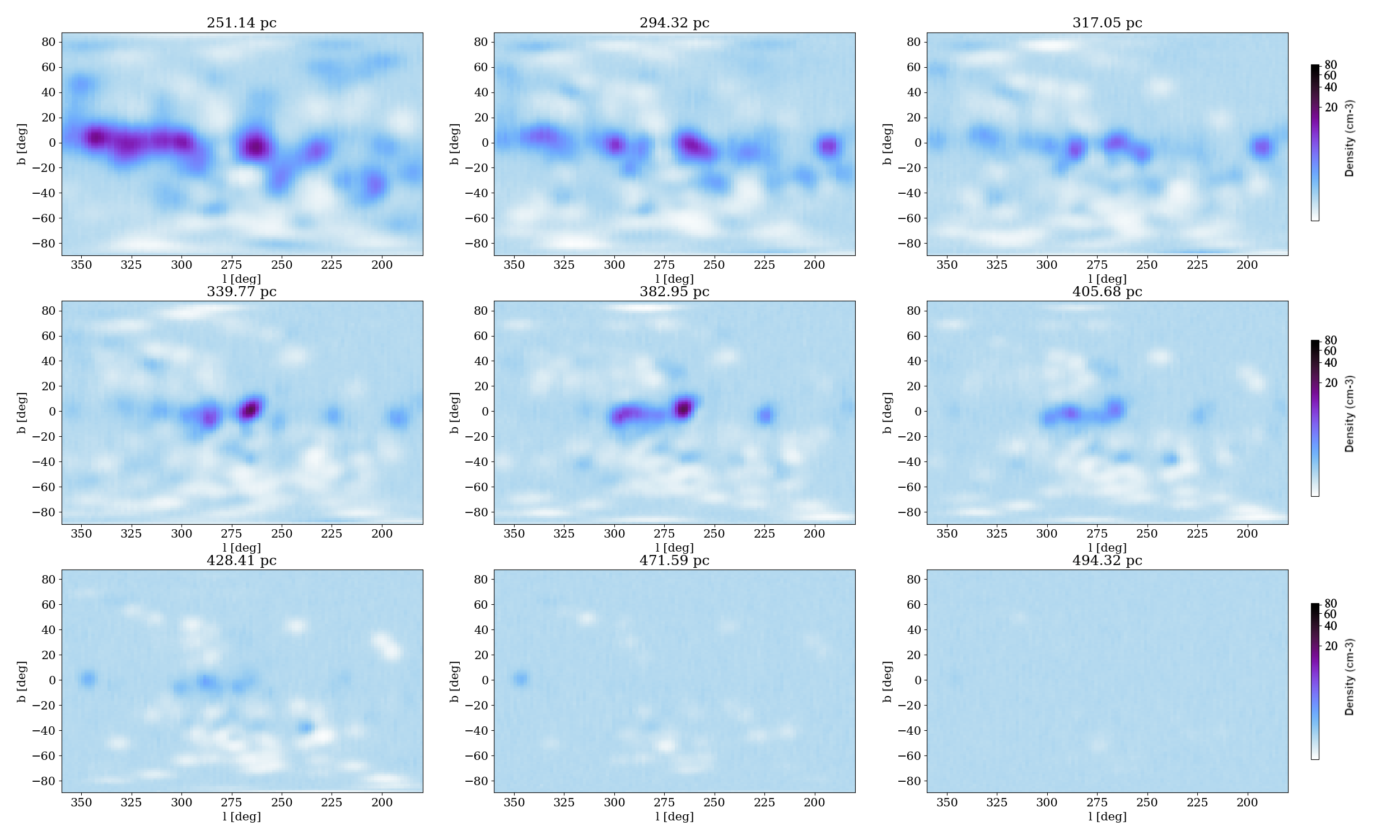} \\
\includegraphics[width=0.98\textwidth]{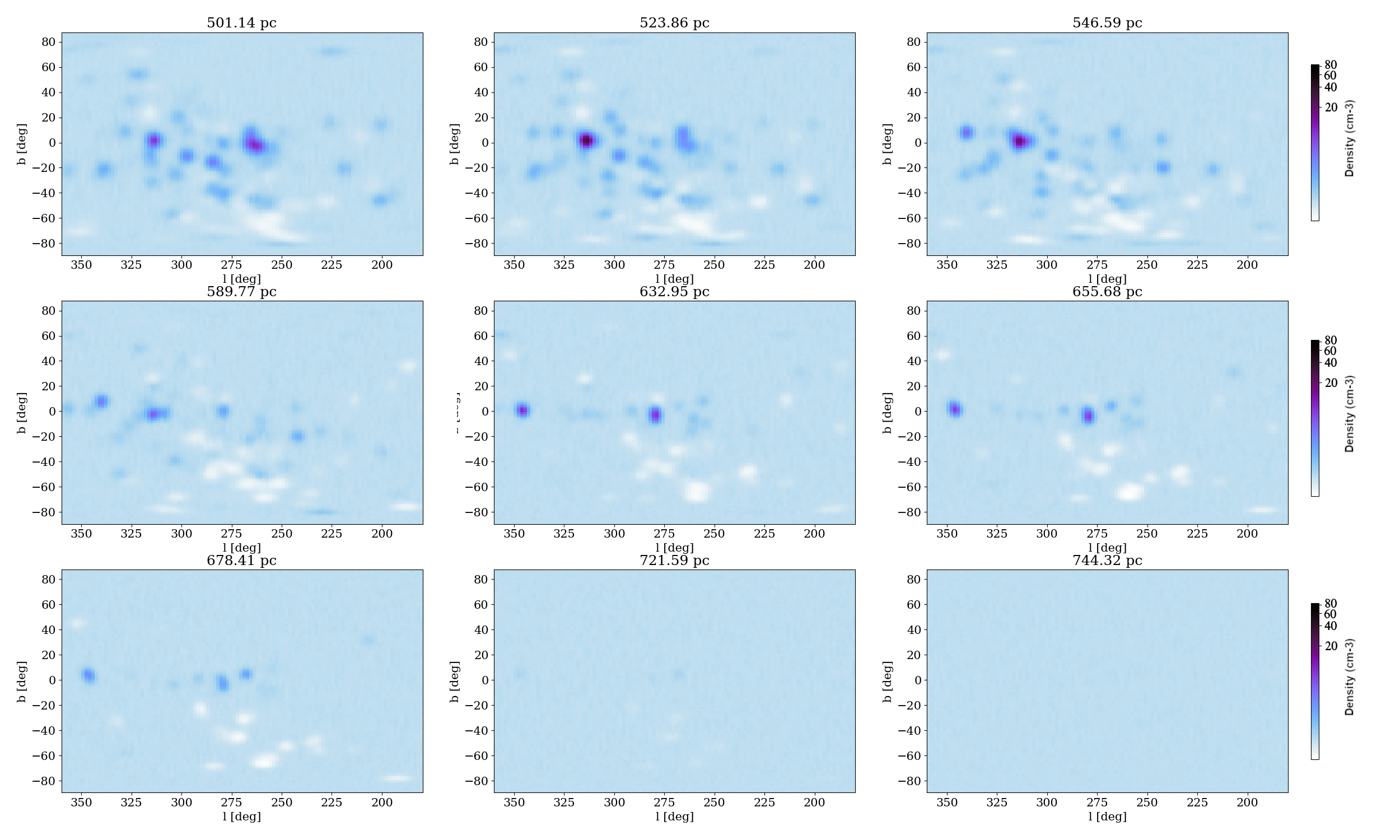} 
      \caption[]{continued
      }\label{fig_3dmap2}
   \end{figure*}

   \begin{figure*}\ContinuedFloat 
\includegraphics[width=0.98\textwidth]{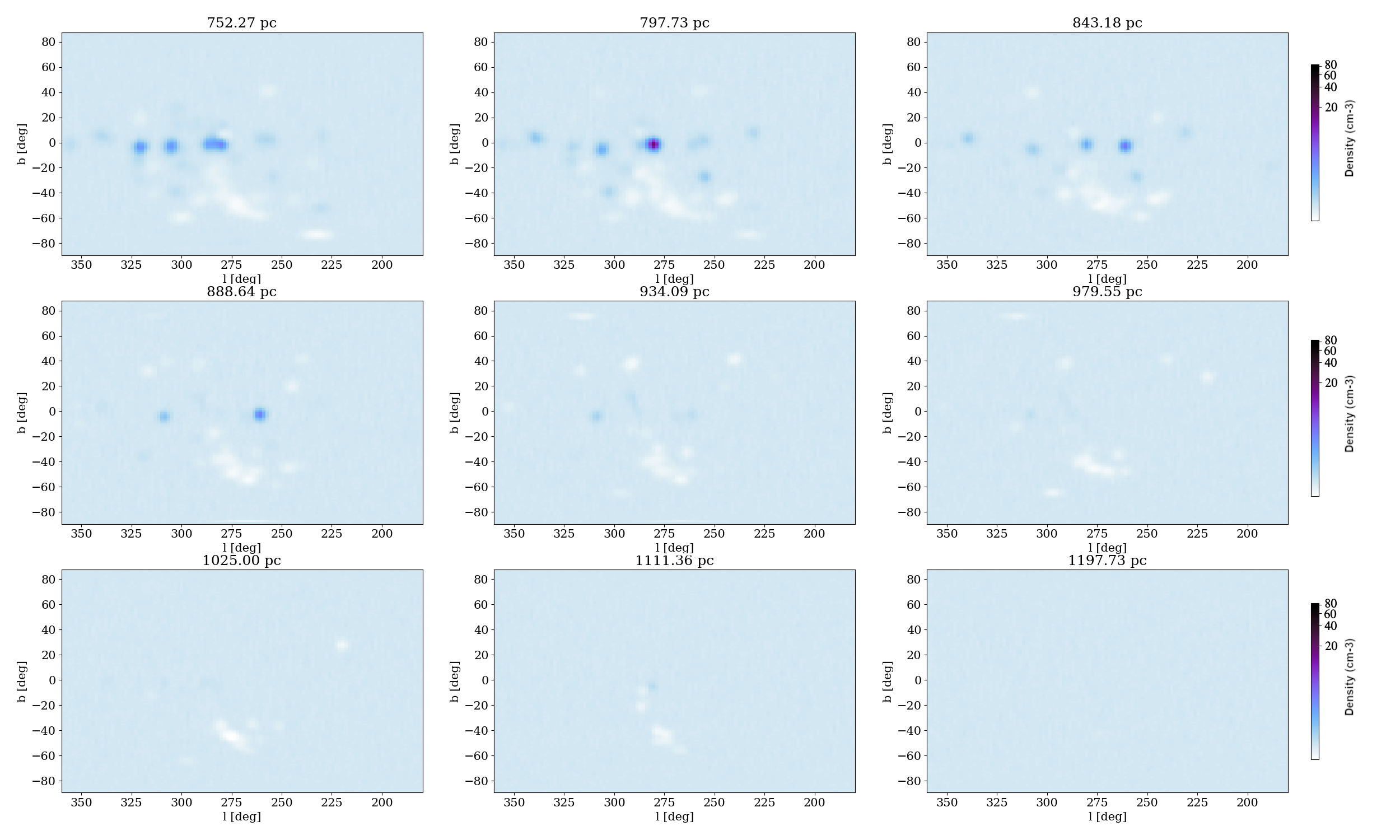}  
      \caption[]{continued
      }\label{fig_3dmap3}
   \end{figure*}

 \section{Summary and Conclusions}\label{sec_con}
Using the X-ray absorption technique, we have performed a detailed study of the hydrogen density distribution in the local ISM.
First, hydrogen column densities were measured by fitting X-ray spectra from coronal sources identified during the inaugural {\it eROSITA} all-sky survey (eRASS1).
Accurate distance measurements were obtained from the {\it Gaia} third data release (DR3) for these sources.
The spectral fitting involved multiple models representing the most commonly observed spectral shapes, with the final column density (NH$_{X}$) selected from the model yielding the best-fit statistic.
Notably, multi-temperature component models were often found to provide the best fit, underscoring the complexity of coronal spectra modeling. 

Our analysis encompassed 8231 sources, a sample size larger than previous studies, covering distances up to 4~kpc. 
Surprisingly, no apparent correlation emerged between column densities and distances or Galactic longitude. 
However, a robust correlation with Galactic latitude suggested a decrease in interstellar material density along the vertical axis moving away from the Galactic plane. 
A comparison with 21~cm measurements revealed instances of high-density points where NH$_{X}$ values tended to be lower than NH$_{21cm}$, potentially influenced by factors like differing effective beam sizes or the presence of absorbers beyond atomic hydrogen.

Further, we compared column densities and extinction values obtained from the SDSS red filter (6231\AA) along various lines of sight.
We computed a slope of $M=0.46\pm 0.02$ between both quantities, in good agreement with previous X-ray absorption analysis but larger than Galactic extinction studies.
Compared with the last ones, it is important to note that our sample is larger, covers high Galactic latitudes, and does not include SNRs, XBs, and PNe, thus avoiding intrinsic absorption, which may affect the measurements.
While no clear indication of variations along the Galactic longitude for the NH$_{X}$/A$_{V}$ ratio was observed, the value decreased moving toward the Galactic plane.

Using equivalent column densities derived from X-ray fits, we modeled the neutral gas distribution with multiple density laws, providing a range of height scale values ($8 < h_{z} < 30$~pc). Unfortunately, due to the absence of sources near the Galactic center, radial scales and central density remained unconstrained. 
Subsequently, employing the hierarchical approach elucidated in \citet{dhar22}, we inferred the hydrogen density distribution from the hydrogen column densities to generate a 3D map of neutral gas in the ISM. 
This approach utilized a Gaussian process to model the $\log_{10}(\rho)$ distribution, with hyperparameters including physical scale lengths in Cartesian coordinates, mean density, and an exponential scale factor. 

Our findings revealed the presence of multiple beams and clouds of varying sizes, indicating the existence of small-scale structures. 
Large density regions around 100~pc, previously identified in dust reddening studies, may be associated with the Galactic Perseus arm. 
Additionally, high-density regions proximal to the Orion star-forming region and the Chamaeleon molecular complex were identified.
This work underscores the capability of the X-ray absorption technique in unveiling the morphological features of the local ISM.

\begin{acknowledgements} 
This work is based on data from {\it eROSITA}, the soft X-ray instrument aboard SRG, a joint Russian-German science mission supported by the Russian Space Agency (Roskosmos), in the interests of the Russian Academy of Sciences represented by its Space Research Institute (IKI), and the Deutsches Zentrum f\"ur Luft- und Raumfahrt (DLR). 
The SRG spacecraft was built by Lavochkin Association (NPOL) and its subcontractors, and is operated by NPOL with support from the Max Planck Institute for Extraterrestrial Physics (MPE). 
The development and construction of the {\it eROSITA} X-ray instrument was led by MPE, with contributions from the Dr. Karl Remeis Observatory Bamberg \& ECAP (FAU Erlangen-Nuernberg), the University of Hamburg Observatory, the Leibniz Institute for Astrophysics Potsdam (AIP), and the Institute for Astronomy and Astrophysics of the University of T\"ubingen, with the support of DLR and the Max Planck Society. 
The Argelander Institute for Astronomy of the University of Bonn and the Ludwig Maximilians Universit\"at Munich also participated in the science preparation for {\it eROSITA}.  
This research was carried out on the High Performance Computing resources of the cobra cluster at the Max Planck Computing and Data Facility (MPCDF) in Garching operated by the Max Planck Society (MPG). 
The {\it eROSITA} data shown here were processed using the eSASS software system developed by the German {\it eROSITA} consortium.   
PCS acknowledges support from DLR through grant 50OR2102.
\end{acknowledgements}

%
%
\bibliographystyle{aa}

\end{document}